\documentstyle[12pt]{article}

\newfont{\twelvemsb}{msbm10 scaled\magstep1}
\newfont{\eightmsb}{msbm8}
\newfam\msbfam
\textfont\msbfam=\twelvemsb
\scriptfont\msbfam=\eightmsb
\catcode`\@=11
\def\Bbb{\ifmmode\let\next\Bbb@\else
  \def\next{\errmessage{Use \string\Bbb\space only in math mode}}\fi\next}
\def\Bbb@#1{{\fam\msbfam{{#1}}}}

\newtheorem{definition}{Definition}
\newtheorem{theorem}{Theorem}
\newtheorem{corollary}{Corollary}

\newcommand{\be}{\begin{equation}}
\newcommand{\ee}{\end{equation}}
\newcommand{\ba}{\begin{eqnarray}}
\newcommand{\ea}{\end{eqnarray}}
\newcommand{\spz}{\hspace{0.5cm}}
\newcommand{\virg}{\spz,\spz}

\newcommand{\finproof}{{\hfill \rule{5pt}{5pt}}}

\def\d{\delta}

\newcommand{\la}{\lambda}

\newcommand{\tp}{\otimes}
\newcommand{\ptp}{\stackrel{\otimes}{,}}
\newcommand{\nn}{\nonumber}

\newcommand{\lqu}{\left[}
\newcommand{\rqu}{\right]}

\newcommand{\buno}{\mbox{\bf 1}}

\newcommand{\cW}{{\cal W}}

\newcommand{\el}{{\mbox {l}}}
\newcommand{\m}{{\mbox {m}}}

\newcommand{\NP}[1]{Nucl.\ Phys.\ {\bf #1}}
\newcommand{\PL}[1]{Phys.\ Lett.\ {\bf #1}}

\newcommand{\CMP}[1]{Comm.\ Math.\ Phys.\ {\bf #1}}
\newcommand{\CPAM}[1]{Comm.\ Pure\ Appl.\ Math.\ {\bf #1}}
\newcommand{\PR}[1]{Phys.\ Rev.\ {\bf #1}}
\newcommand{\PRL}[1]{Phys.\ Rev.\ Lett.\ {\bf #1}}
\newcommand{\MPL}[1]{Mod.\ Phys.\ Lett.\ {\bf #1}}
\newcommand{\IJMP}[1]{Int.\ J.\ Mod.\ Phys.\ {\bf #1}}

\newcommand{\JL}[1]{JETP\ Lett.\ {\bf #1}}
\newcommand{\TMP}[1]{Teor.\ Math.\ Phys.\ {\bf #1}}
\newcommand{\LMP}[1]{Lett.\ Math.\ Phys.\ {\bf #1}}
\newcommand{\FAA}[1]{Funct.\ Anal.\ Appl.\ {\bf #1}}
\newcommand{\JSM}[1]{J.\ Sov.\ Math.\ {\bf #1}}

\begin{document}
\sloppy
\renewcommand{\thefootnote}{\fnsymbol{footnote}}

\newpage
\setcounter{page}{1}

\vspace{0.7cm}
\begin{flushright}
DCPT/01/31 \\
29/2001/EP \\
HWM 01-9 \\
EMPG-01-02 \\
\end{flushright}

\vspace*{1cm}

\begin{center}
{\bf A braided Yang-Baxter Algebra in a Theory of two coupled Lattice Quantum
KdV: algebraic properties and ABA representations.}\\
\vspace{1.8cm}
{\large D.\ Fioravanti $^a$ and M.\ Rossi $^b$ \footnote{E-mail:
Davide.Fioravanti@durham.ac.uk, M.Rossi@ma.hw.ac.uk}}\\ \vspace{.5cm}
$^a${\em Department of Mathematical Sciences, University of Durham, United
Kingdom} \\ and \\ {\em S.I.S.S.A. and Istituto Nazionale di Fisica Nucleare,
Trieste, Italy} \\
\vspace{.3cm}
$^b${\em Department of Mathematics, Heriot-Watt University, Edinburgh,
United Kingdom} \\
\end{center}
\vspace{1cm}

\renewcommand{\thefootnote}{\arabic{footnote}}
\setcounter{footnote}{0}

\begin{abstract}
{\noindent A generalization} of the Yang-Baxter algebra is found in
quantizing the monodromy matrix of two (m)KdV equations discretized
on a space lattice. This braided Yang-Baxter equation still ensures that the
transfer matrix generates operators in involution which form the Cartan
sub-algebra of the braided quantum group. Representations diagonalizing these
operators are described through relying on an easy generalization of Algebraic
Bethe Ansatz techniques. The conjecture that this monodromy matrix algebra
leads, {\it in the cylinder continuum limit}, to a Perturbed Minimal Conformal
Field Theory description is analysed and supported.
\end{abstract}

\vspace{1cm}


\newpage

\section{Introduction.}
After Liouville, an Integrable System (with infinite degrees of freedom) is
usually defined to be a $1+1$ dimensional classical (or quantum) field theory
with the property of having an infinite number of {\it Integrals of Motion in
Involution} ({\bf IMI}). Among them one may be chosen and called the
hamiltonian (operator). As for Quantum Systems the IMI do not help the
determination of the most intriguing and interesting features of these systems
because of their abelian character. However, one can single out at least two
different starting points to overcome this difficulty and both make use of
non-abelian algebras and only partially of the abelian one.

One starting point leaves from the classical theory of integrable systems and more
specifically from the Lax pair formulation of non-linear partial differential
equations \cite{Lax}. Usually, the Poisson structure of the Lax
zero-curvature formulation is encoded in a classical
$r$-matrix \cite{Sk,Fa84,FT} which assures the integrability by entering the
Poisson {\it classical Yang-Baxter algebra} for the entries of the {\it
monodromy matrix}. However, a classical Yang-Baxter algebra is the
expression of an algebraic structure deeper than the abelian one \cite{Dr}.
Indeed, at the quantum level a classical Yang-Baxter algebra becomes
a (quantum) Yang-Baxter algebra (\cite{Ba,KBI} and references within) which
is nothing but a definition relation of a quantum group, a
deformation of an usual Lie algebras \cite{Dr1,Ji,FRT}. As for looking at the
representations of the quantum group from the viewpoint of the spectrum of the
hamiltonian operator, a very efficient evolution of the Bethe Ansatz -- the
Algebraic Bethe Ansatz ({\bf ABA}) -- has been founded initially for the
Sine-Gordon field theory \cite{STF} and then developed for many models
(\cite{KBI} and references  within). In other words, an infinite dimensional
non-abelian algebra {\it includes} the abelian algebra and allows us to build
the spectrum of the hamiltonian operator (and of the others IMI) as its
representation in terms of operators on a Hilbert space (sometimes the
hermitian norm on the space is possibly negative, though always
non-degenerate). More recently, it has been possible to write down exact
non-linear equations describing the energy spectrum of (twisted) Sine-Gordon
field theory on a cylinder \cite{DDV,FMQR}.

Another starting point is based on Statistical Field Theory and in particular
on the very important fact that fixed points of the Renormalization Group are
described by Conformal Field Theories ({\bf CFT's}), {\it i.e.} theories where
the correlation functions are covariant under the conformal group \cite{Pol}.
In 2D the conformal algebra is infinite dimensional ({\it the
Gelfand-Fuks-Virasoro algebra} \cite{GF-V}) and the 2D-CFT's are simple
integrable quantum theories enjoying as their own crucial property the
covariance under an infinite dimensional Virasoro symmetry \cite{BPZ}. As for
the integrability {\it \`a la Liouville} the CFT possesses a bigger
${\cW}$-like symmetry and in particular it is invariant under different
infinite dimensional abelian sub-algebras of the latter \cite{SY}. Each of these
abelian sub-algebras is generated by the IMI, which can be
constructed in terms of the Virasoro algebra, the real new ingredient in these
theories since it is a true field and state spectrum generating symmetry.
Indeed, the Verma modules over this algebra turn out to be
reducible because of the occurrence of sub-modules generated over the
so-called {\it singular vectors} \cite{FF}. The factor-module by the maximal
proper sub-module can be endowed with a non-degenerate hermitian Shapovalov
form and the singular vectors are characterized to produce null hermitian
product with all the other vectors. Now, this factor-module is isomorphic to
the Hilbert space of the local fields (or states) in 2D-CFT's and its own
properties lead to a number of very interesting algebraic-geometrical
features such as character expressions, fusion algebras, differential
equations for  correlation functions, etc. (see \cite{ISZ} for a review).
Unfortunately this beautiful picture collapses when one pushes the system away
from criticality by perturbing the original CFT with some relevant local
field: from the infinite dimensional Virasoro symmetry  only the
finite dimensional Poincar\'e sub-algebra survives the perturbation. After
suitable deformations, at least a conformal abelian sub-algebra
survives the perturbation, resulting in the off-critical abelian algebra
\cite{AZ}. As said before, this symmetry does not carry sufficient information
to find the energy spectrum by means of IMI alone, but it constitutes a very
useful help to determine other interesting quantities. For instance,
scattering theory corresponding to off-critical theories is  usually well
known and contains solitons (or kinks), anti-solitons (or anti-kinks)  and a
number of bound states. The mass spectrum and the S-matrix of different
integrable field theories have been known for about a few years \cite{ms-S}.
Despite this on-shell information, the off-shell Quantum Field Theory is much
less developed. In particular, the computation of the corresponding correlation
functions is still an important open problem. Actually, some progresses
towards this direction have been made, since the exact Form-Factors (FF's) of
several local fields were computed (see for instance \cite{SML,AlZ}). This
allows one to make predictions about the long-distance behavior of the
corresponding correlation functions. On the other hand, some efforts have been
made to estimate the short distance behavior of the theory in the context of
the so-called Conformal Perturbation Theory (CPT) \cite{AlZ}. By combining the
previous techniques (FF's and CPT), it has been possible to estimate several
interesting physical quantities (\cite{FMS} and references therein). In
addition and in the direction of determining in an approximative way the first
energy levels of the simplest perturbed minimal conformal field theories on a
cylinder, very good results have been obtained by the Truncated Conformal Space
Technique, developed in \cite{YZ}. From those results the plane geometry can
be recovered as the limit of cylinder size goes to infinity, on condition of
having a good numerical estimate for large size, which is not so easy to be
obtained.

Consequently, one important problem in Perturbed Conformal Field
Theories ({\bf PCFT's}), {\it i.e.} theories formulated following the second
starting point, is the exact construction of the spectrum of the hamiltonian
operator -- and possibly of the other IMI -- in the more general situation of
the cylinder geometry, by using the idea of the first approach (ABA). This {\it
synergetic} combination of both the previous approaches is difficult in many
cases, {\it i.e.} in all the cases where a Lax formulation of the classical
version of the off-critical theory is missing. Actually, even a quantum Lax
formulation of CFT's is only partially presented and disentangled in the
literature \cite{KM,BLZ,FRS}.

Among the huge variety of integrable theories of the aforementioned kind, the
prototype is the very interesting case of minimal conformal field theories
\cite{BPZ} perturbed by the $\Phi_{1,3}$ primary operator \cite{AZ}. In this
article, a (regularized) lattice integrable definition of the quantum Lax
operator will be given both for the CFT and for the off-critical theory.
Besides, a deep analysis of its algebraic and integrable properties will be
carried out to disentangle the algebraic structure behind the integrability of
the monodromy matrix and of the transfer matrix: a generalization of the
Yang-Baxter equation will be found. In conclusion, a suitable modification of
the ABA will be applied to determine the eigenvalues and eigenstates of the
lattice transfer matrix, the {\it generating function} of all the IMI.
Actually, all the other integrable perturbations of minimal conformal field
theories would be exhausted by treating analogously the conformal case
described in \cite{FRS}, but we will leave this completion for a forthcoming
paper \cite{35}.

In Section 2 we present a brief introduction to classical ($A_1^{(1)}$
modified) KdV theory from the point of view of Lax pair and CFT. In
particular, we show how the space discretization of the monodromy matrix arises
in a very natural way. In Section 3, we look at {CFT} as quantization of
the KdV theory and then propose two left and right lattice regularized
quantum Lax operators. We also calculate explicitly the exchange relations
satisfied by these Lax operators on different sites. In Section 4 we give a
general theorem about the exchange relation satisfied by a general succession
of left and right Lax operators: the conclusion is that in any case we end up
with a braided Yang-Baxter algebra, still ensuring Liouville integrability. In
addition, we single out two {\it conformal} monodromy matrices and two {\it
off-critical} monodromy matrices. In Section 5 we set up the first step towards
the generalization of Algebraic Bethe Ansatz method to braided Yang-Baxter
algebras: the coordinate representation of the basic entries of the
lattice Lax operator. In Section 6 we perform the generalized ABA in the case
of conformal monodromy matrices finding explicitly Bethe Equations and
transfer matrix eigenvalues/eigenvectors. We argue about the insights that
these monodromy matrices describe in the continuum limit the chiral and
anti-chiral part of the minimal CFT's on a cylinder. In Section 7 we
perform the ABA in the case of off-critical monodromy matrices finding
explicitly Bethe Equations and transfer matrix eigenvalues/eigenvectors. In
Section 8 we analyze the {\it conformal} limit on the off-critical transfer
matrices eigenvalues. In Section 9 we disentangle the structure of the
critical and off-critical monodromy matrices in the operatorial scaling limit
to gain understanding about the physical meaning of these theories: in the off-critical
case we guess again that they are equivalent monodromy matrix
descriptions of minimal CFT's perturbed by the $\Phi_{1,3}$ operator. In
Section 10 we find a connection between our braided ABA results and those of
usual ABA in Lattice Sine-Gordon Theory (LSGT). In Section 11 we summarize
our results and give hints about next investigations.

\section{An introduction to the ($A_1^{(1)}$ modified) KdV Theory.}
\setcounter{equation}{0}
It is well known from \cite{SY,KM} that the conformal field theory symmetry
algebra,
\ba
U(y)&=&-\frac{c}{24}+\sum_{-\infty}^{+\infty} L_{-n} e^{iny},  \\
\lqu L_m , L_n \rqu &=& (m-n) L_{m+n}+ \frac{c}{12} (m^3-m) \delta_{m,-n}\, ,
\ea
becomes the second Poisson structure of the usual KdV hierarchy \cite{DS},
\be
\{u(y),u(z)\}=2[u(y)+u(z)]\d^{\prime}(y-z)+\d^{\prime \prime \prime}(y-z),
\label{IIps}
\ee
in {\it the classical limit} (central charge $c\rightarrow -\infty$), provided
the substitutions
\be
U(y)\rightarrow-\frac{c}{6} u(y) \virg [*,*^{\prime}]\rightarrow
\frac{6\pi}{ic} \{*,*^{\prime}\}
\label{claslim}
\ee
are performed. Besides, it has been also established by Drinfeld and Sokolov
\cite{DS} how generalized {\it modified} KdV hierarchies are built through the
centerless Kac-Moody algebras and how the generalized KdV hierarchies
correspond to inequivalent nodes of the Dynkin diagram. In the case of
$A_1^{(1)}$ Dynkin diagram we have the usual KdV hierarchy. For quantization
reasons, we shall start from the usual {\it modified} KdV equation
\begin{equation}
\partial_\tau v =\frac{3}{2} v^2 v^{\prime} + \frac{1}{4}
v^{\prime \prime \prime},  \label{mkdv}
\end{equation}
which describes the temporal flow for the spatial derivative $v=-\varphi
^\prime$  of a Darboux field defined on a spatial interval $y\in [0,R]$,
recalling the connection to the KdV variable $u(y)$ through the Miura
transformation \cite{Miu}:
\begin{equation}
u(y)=\varphi ^\prime (y)^2-i\varphi ^{''}(y) \, .
\end{equation}
Assuming quasi-periodic boundary conditions on $\varphi$, it verifies by
definition the Poisson bracket
\begin{equation}
\{ \varphi (y), \varphi(y^\prime) \}=-\frac{1}{2}s \left(
\frac{y-y^{\prime}}{R} \right) \, ,
\label{classphi}
\end{equation}
where $s(z)$ is the quasi-periodic extension of
the sign function
\begin{equation} s(z)=2n+1 \quad , \quad n<z<n+1 \quad ,
\quad s(n)=2n \quad , \quad n \in {\Bbb Z} \, .
\end{equation}
As a consequence the mKdV variable $v(y)$ satisfies a non-ultralocal Poisson
bracket
\be
\{ v(y), v(y^\prime) \}= \frac {\partial}{\partial y} {\delta^{(p)}} \left(
y-y^{\prime}\right) \, , \ee
the non-ultralocality being expressed by the derivative of the $R$-periodic
delta function $\delta^{(p)}(y)$. Besides, this Poisson structure implies the second
Poisson structure to the KdV field $u$ (\ref{IIps}), which is still
non-ultralocal.

Now, equation (\ref{mkdv}) can be rewritten as a null
curvature condition:  \begin{equation}
[\partial_\tau -{\mbox {l}}^\prime, \partial _y
-{\el}]=0 \end{equation}
for connections belonging to the $A_1^{(1)}$ loop algebra:
\begin{eqnarray}
{\el}&=&-ivh+(e_0+e_1)\, , \label{lleft} \\
{\el}^\prime &=&\lambda ^2(e_0+e_1-ivh)+\frac {1}{2}[(v^2+iv^\prime)e_0
+ (v^2-iv^\prime)e_1]-\frac {1}{2}\left (i \frac {v^{''}}{2}+iv^3\right)h \, ,
\end{eqnarray} where the generators $e_0$, $e_1$, $h$ are chosen in the
canonical gradation of the loop algebra, {\it i.e.}
\begin{equation}
e_0=\lambda E \quad , \quad e_1=\lambda F \quad , \quad h=H \, ,
\end{equation}
with $E$, $F$, $H$ generators of the $A_1$ Lie algebra:
\begin{equation}
[H,E]=2E \quad , \quad [H,F]=-2F \quad ,\quad [E,F]=H.
\end{equation}
For reason of simplicity we choose to deal with the fundamental representation
of $A_1$:
\begin{equation}
H=\left ( \begin{array}{cc} 1 & 0 \\ 0 & -1 \\ \end{array} \right )
\quad , \quad E=\left ( \begin{array}{cc} 0 & 1 \\ 0 & 0 \\
\end{array}\right ) \quad , \quad F=\left ( \begin{array}{cc} 0 & 0 \\
1 & 0 \\ \end{array}\right )\, .
\end{equation}

A remarkable geometrical interest is attached to the monodromy matrix which
realizes  the parallel transport along the {\it space} and which is the
solution of the boundary value problem:
\begin{eqnarray}
\partial _y\, {\m}(y;\lambda)&=&{\el}(y;\lambda)\, {\m} (y;\lambda)\, , \nonumber \\
{\m} (0;\lambda)&=&{\bf 1}\, .
\end{eqnarray}
After indicating with ${\cal P}$ the path-order product, the formal solution
of the previous equation
\begin{equation}
{\m} (y;\lambda)={\cal P}{\mbox {exp}}^{\int _0^ydy^\prime \,  {\el}(y^\prime,\lambda)}
\end{equation}
allows us to calculate the equal time Poisson brackets between the entries of
the monodromy matrix  \begin{equation}
{\m}(\lambda)\equiv {\m}(R;\lambda)={\cal P}{\mbox {exp}}^{\int _0^Rdy \, {\el}(y,\lambda)} \, ,
\label{classmon}
\end{equation}
provided those of the connection ${\el}$ are known. The result is that the Poisson
brackets between the
entries of the monodromy matrix are fixed by the so-called classical
$r$-matrix in the (classical) Yang-Baxter Poisson bracket equation:
\begin{equation}
\{ {\m}(\lambda)  \ptp  {\m}(\lambda ^\prime) \} = [r(\lambda /\lambda ^\prime ), {\m}(\lambda)\tp {\m}(\lambda ^\prime)] \, .
\label {poiss}
\end{equation}
In our particular case the $r$-matrix is the trigonometric one:
\begin{equation}
r(\lambda)=\frac {\lambda +\lambda ^{-1}}{\lambda - \lambda ^{-1}}\frac {H\otimes H}{2}+\frac {2}{\lambda -\lambda ^{-1}}(E\otimes F+F\otimes E)\, .
\end{equation}
By carrying through the trace on both members of the Poisson brackets (\ref
{poiss}),  we are allowed to conclude that the transfer matrix
\begin{equation}
{\mbox {t}}(\lambda )={\mbox {Tr}} \, \m (\lambda )
\label{classtra}
\end{equation}
Poisson-commutes with itself for different values of the spectral parameter:
\begin{equation}
\{ {\mbox {t}}(\lambda ) , {\mbox {t}} (\lambda ^\prime) \} =0 \, . \label{commtra}
\end{equation}
From this relation, we can say that ${\mbox {t}} (\lambda)$ is
the generating function of the classical IMI by expanding it, for instance, in
powers of $\lambda$. As an important example, we obtain a series of local IMI
$I_{2n-1}^{cl}$ from the asymptotic expansion
\begin{equation}
\lambda \rightarrow \infty \quad , \quad \frac {1}{2\pi}\ln {\mbox {t}}
(\lambda)\asymp \lambda - \sum _{n=1}^\infty c_n \lambda
^{(1-2n)}I_{2n-1}^{cl} \, , \label{intmot}
\end{equation}
where $c_n$ are real coefficients (see for example \cite {BLZ} for their
expression). Property (\ref{commtra}) guarantees the
integrability of the model {\it \` a la} Liouville and all the local IMI are
expressed in terms of $u$: for instance the first ones are
\begin{eqnarray}
I_1^{cl}&=&-\frac {1}{2}\int _0^R dy \,  u(y) \, , \nonumber \\
I_3^{cl}&=&-\frac {1}{8}\int _0^R dy \,  u^2(y)
 \, .
\label{inv}
\end{eqnarray}
The equation of motion corresponding to the choice of $I_3^{cl}$ as hamiltonian
\begin{equation}
\partial_{\tau} v = \{ I_3^{cl} \, , \,  v \}
\end{equation}
is the mKdV equation (\ref {mkdv}) itself.

\medskip

In addition, we can introduce {\it the right version} of the mKdV equation:
\begin{equation}
\partial_{\bar \tau} \bar v =\frac {3}{2}\bar v^2\bar v^\prime +\frac {1}{4}
\bar v^{'''} \, , \label{rmkdv} \end{equation}
where $\bar v=-\bar \varphi ^\prime$ and the right quasi-periodic
Darboux variable,
$\bar \varphi (\bar y)$, $0\leq \bar y \leq R$, satisfies the Poisson bracket
(with a change of sign):
\begin{equation}
\{ \bar \varphi (\bar y), \bar \varphi (\bar y^\prime ) \}=\frac{1}{2} s
\left( \frac{\bar y - \bar y^{\prime}}{R} \right) \, ,
 \label{bar}
\end{equation}
and Poisson commutes with the left variable $\varphi (y)$.
Equation (\ref {rmkdv}) derives as in the left case from a null curvature
condition:
\begin{equation}
[\partial _{\bar \tau} -\bar {\el}^\prime, \partial _y -\bar {\el}]=0
\end{equation}
for right connections:
\begin{eqnarray}
\bar {\el}&=&-i\bar vh+(e_0+e_1)\, , \label{lright} \\
\bar {\el}^\prime&=&\lambda ^2(e_0+e_1-i\bar vh)+\frac {1}{2}[(\bar v^2+i\bar v^\prime)e_0+(\bar v^2-i\bar v^\prime)e_1]-\frac {1}{2}\left (i\frac {\bar v^{''}}{2}+i\bar v^3\right)h \, .
\end{eqnarray}

Formul\ae \, for monodromy and transfer matrices are also analogous to the left
case:
\begin{equation}
\bar \m (\lambda)={\cal P}{\mbox {exp}}^{\int _0^R d\bar y\, \bar {\el}
(\bar y,\lambda)} \quad , \quad
\bar {\mbox {t}}(\lambda )={\mbox {Tr}}\, \bar \m (\lambda ) \, .
\label{rclassmon}
\end{equation}

The Poisson brackets between the entries of the monodromy matrix differ for a
sign from their left counterpart:
\begin{equation}
\{ \bar \m (\lambda) \ptp \bar \m (\lambda ^\prime) \} = -[r(\lambda /\lambda
^\prime),\bar \m (\lambda)\tp \bar \m (\lambda ^\prime)] \, , \label {rpoiss}
\end{equation}
which still implies the Poisson-commutativity for the transfer matrix
\begin{equation}
\{ \bar {\mbox {t}}(\lambda ) , \bar {\mbox {t}}(\lambda ^\prime) \} =0 \, .
\end{equation}
$\bar {\mbox {t}}(\lambda)$ generates in its asymptotic expansion the right
classical local IMI: \begin{equation}
\lambda \rightarrow \infty \quad , \quad \ln \bar {\mbox {t}}(\lambda)\asymp
\lambda - \sum _{n=1}^\infty c_n \lambda ^{(1-2n)}\bar I_{2n-1}^{cl} \, ,
\end{equation}
where the $\bar I_{2n-1}^{cl}$ are given by the expressions for $I_{2n-1}^{cl}$ in
which $\varphi$ has been replaced by $\bar \varphi$. Consequently, the first
IMI are:
\begin{eqnarray}
\bar I_1^{cl}&=&-\frac {1}{2}\int _0^R d\bar y\,  \bar u(\bar y) \, ,
\nonumber \\
\bar I_3^{cl}&=&-\frac {1}{8}\int _0^R d\bar y \,\bar u^2(\bar y) \, ,
\label{rinv}
\end{eqnarray}
where
\begin{equation}
\bar u(\bar y)=\bar \varphi ^\prime (\bar y)^2-i\bar \varphi ^{''}(\bar y)
\end{equation}
is the right KdV variable, related to $\bar \varphi$ via the Miura transformation.

Owing to the opposite sign in (\ref {bar}), the right mKdV equation is obtained
through the right action of $\bar I_3$: \begin{equation}
\partial _t \bar v = \{ \bar v  \, , \, \bar I_3^{cl} \} \, .
\end{equation}

\medskip

A very natural way to quantize a classical theory, in presence of path-ordering
and avoiding the problems of ultraviolet divergences, is to put it on the
lattice and then to quantize the  discretized theory. Of course, in case of an
integrable theory the integrability (expressed in our case by the classical
Yang-Baxter equation (\ref {poiss}) and then by the quantum {\it braided} Yang-Baxter equation) has to be preserved by discretization and by quantization.

Hence, let us divide the interval $[0,R]$ in $2N$ parts and define the
discretized Darboux variables:
\begin{equation}
\varphi _k\equiv \varphi (y_k) \quad , \quad \bar \varphi _k \equiv \bar \varphi
(\bar y_k) \quad , \quad y_k\equiv \bar y_k\equiv k\frac {R}{2N} \quad , \quad
k \in {\Bbb Z} \, .
\end{equation}
As a consequence of (\ref {classphi}, \ref {bar}) they satisfy:
\begin{equation}
\{ \varphi _k \, , \, \varphi _h\}=-{\frac {1}{2}}
  {s}\left (\frac {k-h}{2N}\right) \quad , \quad
\{ \bar \varphi _k \, , \, \bar \varphi _h\}={\frac {1}{2}}
  {s}\left (\frac {k-h}{2N}\right) \quad , \quad \{ \varphi _k \, , \, \bar \varphi _h\}=0 \quad .
\label{cphi2}
\end{equation}
We define again for $m \in {\Bbb Z}$:
\begin{eqnarray}
v^-_m\equiv \frac {1}{2}\left [ (\varphi _{2m-1}-\varphi _{2m+1})+(\varphi
_{2m-2}-\varphi _{2m})-(\bar \varphi _{2m-1}-\bar \varphi _{2m+1})+(\bar
\varphi _{2m-2}-\bar \varphi _{2m})\right]   \label {ccorr1}\\
v^+_m\equiv \frac {1}{2}\left [(\bar \varphi _{2m-1}-\bar \varphi _{2m+1})+
(\bar \varphi _{2m-2}-\bar \varphi_{2m})-(\varphi _{2m-1}-\varphi
_{2m+1})+(\varphi _{2m-2}-\varphi _{2m}) \right] .  \label{ccorr2}
\end{eqnarray}
Note that the fields $v^{\pm} _m$ are periodic, {\it i.e.}
$v^{\pm}_{m+N}=v^{\pm}_m$. As a consequence, we can confine ourselves to
the fields $v^{\pm}_m$ with $1\leq m \leq N$. Note also that the fields
$v_m^{\pm}$ live on a lattice which has half the number of sites of the
lattice on which $\varphi _k$ and $\bar \varphi _k$ live. We will indicate with
\begin{equation}
\Delta = \frac {R}{N}
\end{equation}
the lattice spacing of the $v_m^\pm$'s lattice.

Because of (\ref{cphi2}) the operators $v_m^\pm$ enjoy the following
non-ultralocal Poisson brackets:
\begin{eqnarray}
&&\{v^+_m\, , \, v^+_n\}=\frac {1}{2}(\delta^{(p)} _{m-1,n}-\delta^{(p)}
_{m,n-1}) \, , \label{cvrel1} \\
&&\{v^-_m\, , \, v^-_n\}=-\frac {1}{2}(\delta^{(p)} _{m-1,n}-\delta^{(p)}
_{m,n-1}) \, , \label{cvrel2} \\
&&\{v^+_m\, , \, v^-_n\}=-\frac {1}{2}(\delta^{(p)} _{m-1,n}-2\delta^{(p)}
_{m,n}+\delta^{(p)}_{m,n-1}) \,  \label{cvrel3} \, ,
\end{eqnarray}
where $N$-periodic Kronecker delta is defined by
\be
\delta^{(p)}_{m,n}\equiv1\, {\mbox {if}}\,\, (m-n)\in N{\Bbb
Z},\,\,\,\, \equiv0\,\, {\mbox {otherwise}}.
\ee
Therefore, introducing
\begin{equation}
w^\pm_m=e^{iv^\pm_m} \quad ,  \label{cdefn}
\end{equation}
we define the discrete left and right Lax matrices respectively:
\begin{equation}
l_{m}(\lambda)=\left ( \begin{array}{cc} (w_m^-)^{-1} & \Delta \lambda
w_m^+ \\ \Delta \lambda (w_m^+)^{-1} & w_m^- \\ \end{array} \right )
\, , \quad \bar l_{m}(\lambda)=\left ( \begin{array}{cc} (w_m^+)^{-1} &
\Delta \lambda w_m^- \\ \Delta \lambda (w_m^-)^{-1} & w_m^+ \\
\end{array} \right ) \, ,
\label{clax}
\end{equation}
in terms of which the discretized versions of monodromy matrices
(\ref{classmon}) and (\ref {rclassmon}) are:
\begin{eqnarray}
m (\lambda )&=&l_N(\lambda )l_{N-1}(\lambda) \ldots l_2(\lambda)l_1(\lambda) \, , \label {discclass} \\
\bar m (\lambda )&=&\bar l_N(\lambda )\bar l_{N-1}(\lambda) \ldots \bar l_2(\lambda)\bar l_1(\lambda) \, .
\label {rdiscclass}
\end{eqnarray}

Indeed, in the cylinder limit defined by
\begin{equation}
N \rightarrow \infty \spz {\mbox {and fixed}} \spz R\equiv N\Delta  \quad ,
\label{scaling}
\end{equation}
we obtain the scaling equalities
\begin{eqnarray}
&v^-_m =-\Delta \varphi ^\prime (y_{2m})+O(\Delta ^2)&\, ,\label {clim-}\\
&v^+_m = -\Delta \bar \varphi ^\prime
 (\bar y_{2m})+O(\Delta ^2) & \, ,  \label {clim+}
\end{eqnarray}
from which
\begin{equation}
l_{m}(\lambda)= 1+\Delta \,
  {\el}\left ( m\frac {R}{N},\lambda\right ) +O(\Delta ^2) \quad, \quad
\bar l_{m}(\lambda) =
1+\Delta \, \bar { \el}\left (m\frac {R}{N},\lambda \right)+O(\Delta ^2)
 \, . \label{cscallax}
\end{equation}
Therefore the discretized monodromy matrices in the scaling limit behave as
follows:
\begin{eqnarray}
&&m(\lambda) = \prod _{k=1}^N\left [1+\Delta \,
{\el}\left ( k\frac {R}{N},\lambda\right )+O(\Delta ^2)\right ] \rightarrow
{\cal P}{\mbox {exp}}\int _0^Rdy \,
{\el}\left (y, \lambda \right ) =\m (\lambda )\, , \nonumber \\
&&\bar m (\lambda) = \prod _{k=1}^N\left [1+\Delta
\bar {\el}\left ( k\frac {R}{N},\lambda\right )+O(\Delta ^2)\right ] \rightarrow
{\cal P}{\mbox {exp}}\int _0^Rd\bar y \,
\bar {\el} \left (\bar y, \lambda \right ) =\bar \m (\lambda )\, , \nonumber
\end{eqnarray}
{\it i.e.} they reproduce the monodromy matrices for the left and right KdV
theory.

In the next Section, we will quantize the discretized monodromy matrices (\ref
{discclass}, \ref {rdiscclass}) in order to build quantum versions of the left
and right KdV theories.

\section{Quantum version of the KdV theory.}
\setcounter{equation}{0}

The quantum counterparts of the classical local IMI in the KdV
theory are local IMI in conformal field theories \cite{SY} (after suitable
deformation they are local IMI in minimal CFT's perturbed by
the operator $\Phi_{1,3}$ \cite{AZ,EY}). They are constructed in terms of the
quantizations of the Darboux fields, the Feigin-Fuks left and right bosons
\cite{FF}, which we will indicate with $\phi (y)$ and $\bar \phi (y)$. They are
defined to be operators quasi-periodic in $y$ and $\bar y$ verifying the {\it
canonical} (light-cone) commutation relations:
\begin{equation} [\phi (y) \, , \, \phi
(y^\prime)]=-{\frac {i\pi \beta ^2}{2}}s\left (\frac {y-y^\prime }{R}\right)
\quad , \quad   [\bar \phi (\bar y) \, , \, \bar \phi (\bar y^\prime )]={\frac
{i\pi \beta ^2}{2}}s \left (\frac {\bar y-\bar y^\prime}{R}\right) \quad ,
\label{phi}
\end{equation}
where $\beta ^2$ is a real positive constant, and commuting with each other.
By virtue of quasi-periodicity, the fields $\phi$ and $\bar \phi$ can be
expanded in modes as follows:
\begin{eqnarray}
\phi (y)&=&Q+\frac {2\pi y}{R}P-i\sum _{n\not=0}\frac {a_{-n}}{n}e^{i\frac {2\pi}{R}ny} \, , \label{fi} \\
\bar \phi (\bar y)&=&\bar Q-\frac {2\pi \bar y}{R}\bar P-i\sum _{n\not=0}\frac {\bar a_{-n}}{n}e^{-i\frac {2\pi}{R}n\bar y} \, ,\label{fibar}
\end{eqnarray}
and the commutation relations (\ref{phi}) impose that the left and right modes form
two commuting Heisenberg algebras:
\begin{equation}
[Q,P]=[\bar Q,\bar P]=\frac {i}{2}\beta ^2 \quad , \quad [a_n,a_m]=
[\bar a_n,\bar a_m]=\frac {n}{2}\beta ^2
\delta _{n+m,0} \quad ,
\label {modecomm}
\end{equation}
acting respectively on the left and right space whose tensor product
defines the vector space of a conformal field theory (sometimes
the hermitian norm on the space is possibly negative, though always
non-degenerate). In this way, the operators $\phi$ realize a free field
representation of the Virasoro algebra according to the quantum version of the
Miura transformation, called Feigin-Fuks construction \cite{FF}:  \begin{equation}
U(y)=\beta ^{-2}:{\phi ^\prime
(y)}^2:+i(1-\beta ^{-2})\phi ^{\prime \prime}(y)-\frac {1}{24} \quad ,
\label{quaMiu}
\end{equation}
where the symbol normal ordering $: :$ means, as usual, that modes with bigger
index $n$ must be placed to the right. The central charge of this
representation of the Virasoro algebra is
\begin{equation}
c=13-6(\beta ^2 +\beta ^{-2}) \, .
\end{equation}

A whole hierarchy of commuting quantities are built using densities
polynomials of powers of $U(y)$ and its derivatives and they constitute the
chiral quantum local IMI of CFT's \cite{SY}:
\begin{equation}
I_{2k-1}=\int _0^R dy \, U_{2k}(y) \, .
\label{Inv}
\end{equation}
For example, the first densities are:
\begin{equation}
U_2(y)=-\frac {1}{2}\, U(y) \quad , \quad U_4(y)=-\frac {1}{8} :U^2(y): \quad
 . \label {Inv2}
\end{equation}
Of course, after changing $\phi$
with $\bar \phi$, the same construction holds for the right theory. We can define a right Virasoro algebra (we assume the same
central charge as the left algebra)
\begin{equation} \bar U(\bar y)=\beta
^{-2}:{\bar \phi ^\prime (\bar y)}^2:+i(1-\beta ^{-2})\bar \phi ^{\prime
\prime}(\bar y)-\frac {1}{24} \, ,
\label{rquaMiu}
\end{equation}
in terms of which a right hierarchy of commuting quantities is defined
according to formul\ae \, (\ref {Inv}) and (\ref {Inv2}), by replacing $U$ with
$\bar U$. They constitute the right local IMI of CFT's.

In the classical limit (\ref{claslim}) $\beta \rightarrow 0$ and hence
\begin{equation}
[* , *^{\prime}] \rightarrow i\pi \beta ^{2} \{* , *^{\prime}\} \virg
U(y) \rightarrow \beta ^{-2} u(y) \virg \bar U(\bar y) \rightarrow \beta
^{-2} \bar u(\bar y),
\label{compoi}
\end{equation}
in such a way that (\ref {quaMiu}, \ref {rquaMiu})
become the Miura transformations and the IMI of conformal
field theories reduce to the IMI of the KdV theory. Of course, the quantum
Feigin-Fuks operators $\phi $, $\bar \phi$ reduce to the classical Darboux
fields $\varphi$, $\bar \varphi$ respectively.

\medskip

In a natural way we have approached the problem of defining the quantum
versions of the monodromy matrices  (\ref{classmon}, \ref{rclassmon}), so
that we are in the position of deriving expressions for the transfer matrices
and their eigenvectors and eigenvalues. This corresponds to find and
diagonalize the local IMI and also the non-local IMI \cite{FT,BLZ,FRS,FS1}
of quantum KdV (and this IMI are part of those of CFT \cite{FRS}). Besides,
we notice that the continuum methodology developed in a series of beautiful papers
by Bazhanov, Lukyanov and A.B. Zamolodchikov \cite{BLZ} uses slightly
different monodromy matrices than those to which ours reduce in the cylinder
scaling limit (\ref{scaling}). However, we want to remain faithful to the usual
definition of monodromy matrix even in the non-ultralocal case: we will leave
the analysis of the connections to \cite {BLZ} to another paper \cite{35}. Besides, the
construction of a lattice theory will allow us to get rid of  ultraviolet
divergences problems (this statement is pretty obvious but it will be proved
in the next paper \cite{35}) and to use the Algebraic Bethe Ansatz techniques
to diagonalize the monodromy matrix. For all these reasons our starting point
is the quantization of the classical discretized monodromy matrices (\ref
{discclass}, \ref {rdiscclass}).

Let us start with the left case. The discretized quantum Feigin-Fuks bosons
$\phi _k$, $\bar \phi _k$, $k\in {\Bbb Z}$, satisfy (see (\ref{cphi2},\ref{compoi})):
\begin{equation}
[\phi _k \, , \, \phi _h]=-{\frac {i\pi \beta ^2}{2}}
 s\left (\frac {k-h}{2N}\right) \quad , \quad
[\bar \phi _k \, , \, \bar \phi _h]={\frac {i\pi \beta ^2}{2}}
 s \left (\frac {k-h}{2N}\right) \quad , \quad [\phi _k \, , \, \bar \phi _h] =0 \quad .
\label{phi2}
\end{equation}
We define the lattice variables $V^{\pm}_m$, $m\in {\Bbb Z}$, as quantizations
of the classical ones, $v^{\pm}_m$ (\ref {ccorr1}, \ref{ccorr2}):
\begin{eqnarray}
V^-_m\equiv \frac {1}{2}\left [ (\phi _{2m-1}-\phi _{2m+1})+(\phi _{2m-2}-\phi
_{2m})-(\bar \phi _{2m-1}-\bar \phi _{2m+1})+(\bar
\phi _{2m-2}-\bar \phi _{2m})\right]   \label {Ccorr1}\\
V^+_m\equiv \frac {1}{2}\left [(\bar \phi _{2m-1}-\bar \phi _{2m+1})+
(\bar \phi _{2m-2}-\bar \phi_{2m})-(\phi _{2m-1}-\phi
_{2m+1})+(\phi _{2m-2}-\phi _{2m}) \right] .  \label{Ccorr2}
\end{eqnarray}
They are periodic discrete variables: $V^{\pm}_m=V^{\pm}_{m+N}$. Hence, without
loss of generality, we may again restrict ourselves to consider
only $V^{\pm}_m$ with $1\leq m\leq N$. These operators satisfy the non-ultralocal
commutation relations: ($1\leq m,\, n \leq N$)
\begin{eqnarray} &&[V^+_m\, , \,
V^+_n]=\frac {i\pi \beta ^2}{2}(\delta ^{(p)}_{m-1,n}-\delta ^{(p)}_{m,n-1})
\, , \label{vrel1} \\ &&[V^-_m\, , \, V^-_n]=-\frac {i\pi \beta ^2}{2}(\delta
^{(p)} _{m-1,n}-\delta ^{(p)}_{m,n-1}) \, , \label{vrel2} \\ &&[V^+_m\, , \,
V^-_n]=-\frac {i\pi \beta ^2}{2}(\delta ^{(p)} _{m-1,n}-2\delta
^{(p)}_{m,n}+\delta ^{(p)} _{m,n-1}) \,  \label{vrel3} \, . \end{eqnarray}
Therefore, after defining
\begin{equation}
W^\pm_m \equiv e^{iV^\pm_m} \quad ,
\quad q \equiv e^{-i\pi \beta ^2} \, , \label{defn} \end{equation}
we can derive from the commutator algebra (\ref{vrel1}-\ref{vrel3}) the
exchange algebra:
\begin{eqnarray}
&&W^{\pm}_{m+1}W^{\pm}_m=q^{\pm \frac {1}{2}}W^{\pm}_mW^{\pm}_{m+1} \, , \, \,
W^{\pm}_{m+1}W^{\mp}_m=q^{\mp \frac {1}{2}}W^{\mp}_mW^{\pm}_{m+1} \, ;\nn \\
&&\hspace {2.94cm}
W^+_mW^-_m=qW^-_mW^+_m\, ; \label{Wrel} \\
&& \hspace {2.9cm} [W_m^{\sharp},W_n^{\sharp ^\prime}]=0 \, \,   \quad {\mbox
{if}} \, \, \, (1\leq n \leq N) \, \,2\leq |m-n|\leq N-2 \, ,\nn
\end{eqnarray}
with the obvious identification $W^{\pm}_{N+1}=W^{\pm}_1$ and
with $\sharp$ and $\sharp^\prime$ both equal to $+$ or $-$. Plus or minus part
of this algebra has been introduced in \cite{FV}. At the end, we define the
discrete Lax operators \begin{equation} L_{m}(\lambda) \equiv \left (
\begin{array}{cc} (W_m^-)^{-1} & \Delta \lambda W_m^+ \\ \Delta \lambda
(W_m^+)^{-1} & W_m^- \\ \end{array} \right ) \, , \quad \bar L_{m}(\lambda)
\equiv \left ( \begin{array}{cc} (W_m^+)^{-1} & \Delta \lambda W_m^- \\ \Delta
\lambda (W_m^-)^{-1} & W_m^+ \\ \end{array} \right ) \, ,
\label{lax}
\end{equation}
which are a quantization of the discrete left and right Lax matrices
(\ref{clax}).

Operators $L_m$ were used in \cite{Kun} for defining the discretized
monodromy  matrix of the (left) KdV theory as
\begin{equation}
L_N(\lambda )L_{N-1}(\lambda) \ldots L_2(\lambda)L_1(\lambda)
\label{kundu}.
\end{equation}
As it will be clear in the following, this definition is perfectly correct,
although the ABA solution of the problem in \cite{Kun} contains an {\it
ab initio} mistake which affects the final results (the author of \cite{Kun}
is in agreement with our finding \cite{Kunc}). In addition, we
introduced the right chiral counterpart of $L_{m}$, the Lax operator
$\bar L_{m}$.

Now, it is important for the following to derive the exchange  relations for
left and right Lax operators (\ref {lax}). Hence, let us introduce the
quantum $R$-matrix and the quantum $Z$-matrix, the matrix encoding the braiding:
\begin{equation}
R_{ab}(\xi )=\left ( \begin{array}{cccc} 1 & 0 & 0 & 0 \\
0 & \frac {\xi ^{-1}-\xi}{q^{-1}\xi ^{-1}-q\xi} & \frac
{q^{-1}-q}{q^{-1}\xi ^{-1}-q\xi } & 0 \\
0 & \frac {q^{-1}-q}{q^{-1}\xi ^{-1}-q\xi} &\frac
{\xi ^{-1}-\xi }{q^{-1}\xi ^{-1}-q\xi } & 0 \\ 0 & 0 & 0 & 1 \end{array} \right)
\, ,
\label {Rmat}
\end{equation}
\begin{equation}
Z_{ab}=\left ( \begin{array}{cccc} q^{-\frac {1}{2}} & 0 & 0 & 0 \\
0 & q^{\frac {1}{2}} & 0 & 0 \\ 0 & 0 & q^{\frac {1}{2}} & 0 \\
0 & 0 & 0 & q^{-\frac {1}{2}} \end{array} \right) \, ,
\label{Zmat}
\end{equation}
which act on the tensor product $a\otimes b$ of two auxiliary
two-dimensional spaces. Using only the exchange relations (\ref{Wrel}) one
can show that the operators (\ref{lax}) satisfy the following relations:

\begin{enumerate}

\item For $1\leq m \leq N$:

\begin{eqnarray}
R_{ab}\left (\frac {\lambda}{\lambda ^\prime}\right)L_{am}(\lambda)
L_{bm}(\lambda ^\prime)&=&
L_{bm}(\lambda ^\prime)L_{am}(\lambda)R_{ab}\left(\frac
{\lambda}{\lambda ^\prime} \right) \,
\label {Lrel1},\\
R_{ab}\left (\frac {\lambda ^\prime}{\lambda}\right)\bar L_{a\,m}(\lambda)
\bar L_{b\,m}(\lambda ^\prime)&=&\bar L_{b\,m}(\lambda ^\prime)
\bar L_{a\,m}(\lambda)
R_{ab}\left(\frac {\lambda ^\prime}{\lambda}\right) \,
\label {Lrel2} ;
\end{eqnarray}

\item For $1\leq m\leq N-1$:

\begin{eqnarray}
L_{am}(\lambda)L_{bm+1}(\lambda ^\prime)&=&L_{bm+1}(\lambda ^\prime)
Z_{ab}^{-1}L_{am}(\lambda)
\, , \label {Lrel3}\\
\bar L_{am}(\lambda)\bar L_{bm+1}(\lambda ^\prime)&=&\bar
L_{bm+1}(\lambda ^\prime) Z_{ab}
\bar L_{am}(\lambda) \, ,\label {Lrel4} \\
L_{am}(\lambda )\bar L_{bm+1}(\lambda ^\prime)&=&\bar L_{bm+1}(\lambda
^\prime )Z_{ab}^{-1}L_{am}(\lambda)
\, , \label {Lrel5}\\
\bar L_{am}(\lambda )L_{bm+1}(\lambda ^\prime)&=&L_{bm+1}(\lambda
^\prime )Z_{ab}\bar L_{am}(\lambda)
\, ;
\label {Lrel6}
\end{eqnarray}

\item $N$-$1$ exchange:

\begin{eqnarray}
L_{aN}(\lambda)L_{b1}(\lambda ^\prime)&=&L_{b1}(\lambda ^\prime)
Z_{ab}^{-1}L_{aN}(\lambda)
\, , \label {Lrel7}\\
\bar L_{aN}(\lambda)\bar L_{b1}(\lambda ^\prime)&=&\bar
L_{b1}(\lambda ^\prime) Z_{ab}
\bar L_{aN}(\lambda) \, ,\label {Lrel8} \\
L_{aN}(\lambda )\bar L_{b1}(\lambda ^\prime)&=&\bar L_{b1}(\lambda
^\prime )Z_{ab}^{-1}L_{aN}(\lambda)
\, , \label {Lrel9}\\
\bar L_{aN}(\lambda )L_{b1}(\lambda ^\prime)&=&L_{b1}(\lambda
^\prime )Z_{ab}\bar L_{aN}(\lambda)
\, .
\label {Lrel10}
\end{eqnarray}

\end{enumerate}

In these equations we have defined $L_{am}\equiv L_m(\lambda)\otimes \buno$ and
$L_{bm}\equiv \buno\otimes L_m(\lambda )$. The first two relations are just
Yang-Baxter equations, while the others describe the non-ultralocality, {\it
i.e.} the fact that Lax operators on first-neighboring sites and different auxiliary spaces do not commute. Of course, operators (\ref
{lax}) on different auxiliary spaces and on sites $m$ and $n$ commute if
$\, 2\leq |m-n|\leq N-2$.

In spite of this complication, it has been shown in \cite{Kun} that the
monodromy matrix (\ref{kundu}) satisfies a modified version of the Yang-Baxter
equation, called braided Yang-Baxter equation, and that the corresponding
transfer matrices are commuting operators for different values of the spectral
parameter.

\medskip

\section{Braided Yang-Baxter algebra and Integrals of Motion.}
In this section we will define in a general way monodromy matrices as products
of operators $L$ and $\bar L$ (\ref {lax}) in every possible order. Then we
will prove that every monodromy matrix generates the braided
Yang-Baxter algebra.

Let us introduce the following site operators ($1 \leq m \leq N$):
\begin{equation}
K_{m}(\lambda)\equiv \chi _m L_{m}(\lambda\delta _m)+\bar \chi _m
\bar L_{m}\left (\frac {\delta _m}{\lambda}\right)
\label{Lambda} \, ,
\end{equation}
where, for a fixed $m$, the real numbers $\chi_m$, $\bar \chi _m$ may take
only the two set of values
\begin{equation}
\{ \chi _m =0\, , \, \bar \chi _m=1 \} \quad {\mbox {or}} \quad \{ \chi _m =1
\, , \, \bar \chi _m=0 \} \quad ,
\label{tau}
\end{equation}
whereas $\delta _m$ are arbitrary complex parameters. In other words on a
fixed lattice site $m$ the operator $K_m(\lambda )$ can be equal to
$L_m(\lambda \delta _m)$ or to $\bar L_m(\delta _m/\lambda)$.

By using properties (\ref {Lrel1}-\ref {Lrel10}) and conditions
(\ref {tau}) we can show very easily that:
\begin{enumerate}

\item For $1\leq m \leq N$:

\begin{equation}
R_{ab}\left (\frac {\lambda}{\lambda ^\prime}\right)K
_{am}(\lambda)K_{bm}(\lambda ^\prime)= K _{bm}(\lambda
^\prime) K_{am}(\lambda) R_{ab}\left(\frac {\lambda}{\lambda ^\prime}\right) \,
; \label {Lambdarel1}
\end{equation}

\item For $1\leq m \leq N-1$:

\begin{equation}
K _{am}(\lambda)K_{bm+1}(\lambda ^\prime)=K_{bm+1}(\lambda ^\prime)
[\chi _m Z_{ab}^{-1}+\bar \chi _m Z_{ab}]K_{am}(\lambda )
\, ; \label{Lambdarel2}
\end{equation}

\item $N$-$1$ exchange:
\begin{equation}
K _{aN}(\lambda)K_{b1}(\lambda ^\prime)=K_{b1}(\lambda ^\prime)
[\chi _N Z_{ab}^{-1}+\bar \chi _N Z_{ab}]K_{aN}(\lambda )
\, . \label{Lambdarel3}
\end{equation}
\end{enumerate}

Operators (\ref{Lambda}) on sites $m$ and $n$ and on different auxiliary spaces
commute if $2\leq |m-n|\leq N-2$. Now we are in the position to define in
complete generality the monodromy matrix mentioned at the beginning of this
Section: \begin{equation}
\Pi (\lambda)\equiv K _{N}(\lambda)\ldots K
_{1}(\lambda) \label{mong} \, .
\end{equation}
Thanks to (\ref{Lambda}, \ref{tau}) the matrix (\ref{mong}) is an ordered
product of operators which for a fixed lattice site $m$ may be equal to
$L_m(\lambda \delta _m)$ or to $\bar L_m(\delta _m/\lambda )$. In
particular, the left monodromy matrix (\ref{kundu}) of \cite{Kun} is obtained
when $\chi _m =1 \, , \delta _m =1\, , \forall m$. Besides, The right analogue
of this monodromy matrix is obtained when $\bar \chi _m =1 \, , \delta _m =1 \,
, \forall m$.

Let us now state the key-theorem of this Section.
\begin{theorem}
\label{thmmon}
The monodromy matrix (\ref {mong}) satisfies
for $N\geq2$ the following braided relations:
\begin{equation}
R_{ab}\left ( \frac {\lambda}{\lambda ^\prime}\right)\Pi _a(\lambda)
\left [\chi _NZ_{ab}^{-1}+\bar \chi _NZ_{ab}\right ]\Pi
_b(\lambda ^\prime)=\Pi_b(\lambda ^\prime)\left [\chi _NZ_{ab}^{-1}+\bar \chi _NZ_{ab}\right]
\Pi _a(\lambda)R_{ab}\left(\frac {\lambda}{\lambda ^\prime}
\right) \, .
\label{pmon2}
\end{equation}
\end{theorem}

\noindent
{\bf Proof}: The proof follows by the repeated applications of relations
(\ref {Lambdarel1},\ref {Lambdarel2},\ref {Lambdarel3}). \finproof

\medskip
\begin{definition}
An associative algebra generated by the entries $\Pi _{ij} (\lambda )$ of a
2 by 2 matrix $\Pi (\lambda)$ satisfying the relation:
\begin{equation}
R_{ab}\left ( \frac {\lambda}{\lambda ^\prime}\right){\cal Z}_{ba}^{-1}\Pi _a(\lambda) {\hat {\cal Z}_{ab}}^{-1}\Pi
_b(\lambda ^\prime)={\cal Z}_{ab}^{-1}\Pi_b(\lambda ^\prime)
 {\hat {\cal Z}_{ba}}^{-1}\Pi _a(\lambda)R_{ab}\left(\frac {\lambda}{\lambda ^\prime}
\right) \, ,
\label{zyb}
\end{equation}
where $R_{ab}(\xi)$, ${\cal Z}_{ab}$ and ${\hat {\cal Z}_{ab}}$ are 4 by 4 numerical matrices obeying:
\begin{eqnarray}
R_{ab}(\xi)R_{ac}(\xi \xi ^\prime)R_{bc}(\xi ^\prime)&=&R_{bc}(\xi ^\prime)
R_{ac}(\xi \xi ^\prime)R_{ab}(\xi) \label {yb}\\
{\cal Z}_{ab}{\cal Z}_{ac}{\cal Z}_{bc}&=&{\cal Z}_{bc}{\cal Z}_{ac}{\cal Z}_{ab} \label {rz1} \\
\hat {\cal Z}_{ab}\hat {\cal Z}_{ac}{\cal Z}_{bc}&=&{\cal Z}_{bc}\hat {\cal Z}_{ac}\hat {\cal Z}_{ab} \label {rz2}\\
R_{ba}(\xi)\hat {\cal Z}_{ac}\hat {\cal Z}_{bc}&=&\hat {\cal Z}_{bc}\hat {\cal Z}_{ac}R_{ba}(\xi) \label {rz3} \\
R_{cb}(\xi)\hat {\cal Z}_{ac}\hat {\cal Z}_{ab}&=&\hat {\cal Z}_{ab}\hat {\cal Z}_{ac}R_{cb}(\xi) \label {rz4}\\
R_{ba}(\xi) {\cal Z}_{ac}{\cal Z}_{bc}&=& {\cal Z}_{bc} {\cal Z}_{ac}R_{ba}(\xi) \label {rz5}\\
R_{cb}(\xi){\cal Z}_{ac}{\cal Z}_{ab}&=& {\cal Z}_{ab} {\cal Z}_{ac}R_{cb}(\xi)
\quad , \label {rz6}
\end{eqnarray}
is called braided Yang-Baxter algebra. Equation (\ref {zyb}) is called braided Yang-Baxter equation.
\end{definition}

Braided Yang-Baxter algebras have been introduced in \cite {HLA}.

Equations (\ref {yb}-\ref {rz6}) guarantee the associativity of the triple product:
\begin{equation}
\Pi _a (\lambda ){\hat {\cal Z}_{ab}}^{-1}\Pi
_b(\lambda ^\prime){\hat {\cal Z}_{ac}}^{-1}{\hat {\cal Z}_{bc}}^{-1}
\Pi _c (\lambda ^{''}) \nonumber \, .
\end{equation}

In our specific case $R_{ab}$ is given by (\ref {Rmat}), while:
\begin{equation}
{\cal Z}_{ab}={\hat {\cal Z}_{ab}}=[\chi _NZ_{ab}+\bar \chi
_NZ_{ab}^{-1}] \, . \nonumber
\end{equation}
Since:
\begin{equation}
[R_{ab}(\xi), Z_{ab}]=0 \, ,
\label {RZ}
\end{equation}
relation (\ref {zyb}) reduces to (\ref {pmon2}).

Matrix (\ref {Rmat}) is well known to satisfy Yang-Baxter equation (\ref {yb})
and from (\ref {RZ}) and the fact that $Z_{ab}$ is diagonal the other
associativity conditions (\ref {rz1}-\ref {rz6}) follow straightforwardly.

The braided Yang-Baxter algebra is a generalization of the usual
Yang-Baxter algebra in the sense that in the particular case ${\cal Z}_{ab}=\hat {\cal Z}_{ab}=\buno$ the former reduces to the latter. In our particular case, after
looking at the form of $Z_{ab}$ (\ref{Zmat}), we can say that this may occur
only for the special value of the deforming parameter $q=1$: this is why we
call this algebra a braided generalization of Yang-Baxter algebra rather than a
deformed generalization.

We also observe that a simple consequence of Theorem
1 is that there is no way to reproduce Yang-Baxter algebra by {\it fusing}
site Lax operators (\ref{Lambda}): therefore the presence of the braided
Yang-Baxter equation is an unavoidable feature of our approach, which, in its
turn, leaves very naturally from the algebraic formulation of CFT's.

As a corollary of the previous theorem, we now prove the Liouville
integrability.
\begin{corollary}
\label {cortra}
The $\lambda$-dependent {\it transfer matrix}
\be
\sigma(\la) \equiv {\mbox {Tr}}\, \Pi (\lambda)\,
\label{transf}
\end{equation}
commutes with itself at different values of $\la$:
\begin{equation}
\left [ {\mbox {Tr}} \, \Pi (\lambda) \, \, , \,   {\mbox {Tr}}\, \Pi
(\lambda ^\prime )\right ] = 0 \, .
\label{comm2}
\end{equation}
\end{corollary}

\noindent
{\bf Proof}: After multiplying relation (\ref {pmon2}) by $\chi _NZ_{ab}+\bar
\chi _NZ_{ab}^{-1}$ and after using the aforementioned property:
\begin{equation}
[R_{ab}(\la),Z_{ab}]=0 \, , \nonumber
\end{equation}
we obtain
\begin{eqnarray}
&&\left [\chi _NZ_{ab}+\bar \chi
_NZ_{ab}^{-1}\right ]\Pi _a(\lambda)\left [\chi _NZ_{ab}^{-1}+\bar \chi
_NZ_{ab}\right ]\Pi
_b (\lambda ^\prime)=\nonumber \\
&&=R_{ab}\left (\frac {\lambda }{\lambda^\prime}\right)^{-1}\left [\chi
_NZ_{ab}+\bar \chi _NZ_{ab}^{-1}\right ]\Pi_b(\lambda ^\prime)
\left [\chi _NZ_{ab}^{-1}+\bar \chi _NZ_{ab}\right ]\Pi _a(\lambda)
R_{ab}\left(\frac {\lambda}{\lambda ^\prime} \right) \, \nonumber .
\end{eqnarray}
Then, from the cyclicity of the trace, we have
\begin{eqnarray}
&& {\mbox {Tr}}_{ab}\left \{ \left [\chi _NZ_{ab}+\bar \chi
_NZ_{ab}^{-1}\right]\Pi _a(\lambda)\left[\chi _NZ_{ab}^{-1}+\bar \chi _N
Z_{ab}\right]\Pi _b(\lambda ^\prime)\right \}= \nonumber \\
&&= {\mbox {Tr}}_{ab}\left \{ \left[\chi _NZ_{ab}+\bar \chi
_NZ_{ab}^{-1}\right]\Pi _b(\lambda ^\prime)
\left[\chi _NZ_{ab}^{-1}+\bar \chi
_NZ_{ab}\right]\Pi _a (\lambda)\right \} \, .
\label{prop2}
\end{eqnarray}
From the diagonal structure of $Z$ we can write explicitly
\begin{equation}
[\chi _NZ+\bar \chi_NZ^{-1}]_{\alpha _1 \alpha _2}^{\beta _1 \beta _2}=z_{\alpha _1 \alpha _2}\, \delta _{\alpha _1}^{\beta _1}\delta _{\alpha _2}^{\beta _2}
\quad , \quad [\chi _NZ^{-1}+\bar \chi_NZ]_{\alpha _1 \alpha _2}^{\beta _1 \beta _2}=z_{\alpha _1 \alpha _2}^{-1}\, \delta _{\alpha _1}^{\beta _1}\delta _{\alpha _2}^{\beta _2} \, , \nonumber
\end{equation}
where $z_{\alpha _1 \alpha _2}$ are some complex numbers. Hence,
the property (\ref{prop2}) can be re-written explicitly as
\begin{equation}
\sum _{\alpha _1\, ,  \alpha _2}z_{\alpha _1 \alpha _2} \Pi (\lambda )_{\alpha _1}^{\alpha _1}\, z_{\alpha _1 \alpha _2}^{-1} \Pi (\lambda ^\prime)_{\alpha _2}^{\alpha _2}=\sum _{\alpha _1\, ,  \alpha _2}z_{\alpha _1 \alpha _2} \Pi (\lambda ^\prime)_{\alpha _2}^{\alpha _2}\, z_{\alpha _1 \alpha _2}^{-1} \Pi (\lambda )_{\alpha _1}^{\alpha _1} \, ,
\nonumber
\end{equation}
which shows the commutativity of the transfer matrices
${\mbox {Tr}}\, \Pi (\lambda)$ for different values of the spectral parameter
$\la$.

\medskip

At the end of this Section, we define some important examples of monodromy
matrices which we will deal with.

\begin{itemize}

\item CONFORMAL CASE:
\begin{enumerate}
\item Left Monodromy Matrix
\ba
&& \quad \chi _m=1 \quad , \quad \bar \chi _m= 0 \quad , \quad \delta _m=1 \Rightarrow \nonumber \\
&& \Rightarrow \Pi(\lambda) = M(\lambda) \equiv L_{N}(\lambda )
\ldots L_{1}(\lambda );
\label{leftmon}
\ea
\item Right Monodromy Matrix
\ba
&&  \quad \chi _m=0 \quad , \quad  \bar \chi _m=1 \quad , \quad \delta _m=1
\Rightarrow \nonumber \\ &&\Rightarrow \Pi (\lambda) = \bar M
(\lambda) \equiv \bar L_{N}\left (\frac {1}{\lambda } \right)\ldots \bar L_{1}
\left (\frac {1}{\lambda } \right) .
\label{rightmon}
\ea
\end{enumerate}
\end{itemize}

\begin{itemize}

\item OFF-CRITICAL CASE:
\begin{enumerate}
\item Case right-left (r-l)
\ba
&& \quad \chi _{4i}=\chi
_{4i-1}=0 \quad , \quad \bar \chi _{4i-2}=\bar \chi _{4i-3}=0 \quad \left
(1\leq i \leq \frac {N}{4}\, , \, \, N \in 4 {\Bbb N}\right )\, ,  \nonumber
\\ && \delta _{m}=\mu^{\frac {1}{2}} \quad (1\leq m\leq N ) \Rightarrow
\nonumber \\ && \Rightarrow \Pi (\lambda) = {\bf M}(\lambda) \equiv \bar
L_{N}\left (\frac {\mu^{\frac {1}{2}}}{\lambda}\right) \bar L_{N-1}\left
(\frac {\mu^{\frac {1}{2}}}{\lambda}\right)L_{N-2}(\lambda \mu ^{\frac
{1}{2}})L_{N-3}(\lambda \mu ^{\frac {1}{2}}) \ldots \nonumber \\  && \ldots
\bar L_{4}\left (\frac {\mu^{\frac {1}{2}}}{\lambda}\right) \bar L_{3}\left
(\frac {\mu^{\frac {1}{2}}}{\lambda}\right)L_{2}(\lambda \mu ^{\frac
{1}{2}})L_{1}(\lambda \mu^{\frac {1}{2}}) \, ;
\label {Rightmon}
\ea
\item Case left-right (l-r)
\ba
&& \quad \bar \chi _{4i}=\bar \chi _{4i-1}=0 \quad , \quad \chi _{4i-2}=\chi
_{4i-3}=0 \quad \left (1\leq i \leq \frac {N}{4}\, , \, \, N \in 4
{\Bbb N}\right )\, , \nonumber \\
&& \delta _{m}=\mu ^{\frac {1}{2}} \quad (1\leq m\leq N ) \Rightarrow
\nonumber \\ && \Rightarrow \Pi (\lambda) = {\bf M}^\prime (\lambda) \equiv
L_{N}\left ({\lambda}\mu^{\frac {1}{2}}\right) L_{N-1}\left
({\lambda}\mu^{\frac {1}{2}}\right)\bar L_{N-2}\left (\frac  {\mu ^{\frac
{1}{2}}}{\lambda} \right)\bar L_{N-3}\left (\frac {\mu ^{\frac
{1}{2}}}{\lambda} \right) \ldots \nonumber \\  && \ldots  L_{4}({\lambda} \mu^{\frac {1}{2}}) L_{3} ({\lambda}\mu^{\frac {1}{2}})  \bar L_{2}\left (\frac
{\mu ^{\frac {1}{2}}}{\lambda}\right) \bar L_{1}\left (\frac {\mu ^{\frac
{1}{2}}}{\lambda} \right) \, .  \label {Leftmon}  \ea
\end{enumerate}

\end{itemize}

Now, we must give some explanation about names which have a physical origin.
The monodromy matrix (\ref{leftmon}) has been introduced as a natural
discretized version of that describing Quantum KdV Theory, {\it i.e.} the
left part of CFT \cite{BLZ}. The monodromy matrix
(\ref{rightmon}) is  simply its right counterpart, completing the description
of CFT. The quantum KdV description of CFT exhibits, for
particular values of $\beta^2$, the usual features of conformal minimal CFT's
perturbed by the $\Phi_{1,3}$ operator ({\it e.g.} the form of local IMI) \cite{FRS}. Hence, this formulation should be very suitable for
going into the off-critical region preserving integrability and our
proposal (\ref {Rightmon},\ref {Leftmon}) for the description of CFT's
perturbed by the $\Phi_{1,3}$ operator is now very natural. In any case, we
will bring other supports to our conjecture in the following by diagonalizing
the transfer matrices corresponding to (\ref{leftmon}-\ref{Leftmon}) through
ABA techniques.

\section{Coordinate representation.}
\setcounter{equation}{0}
In order to settle down a suitable generalization of ABA to the braided
Yang-Baxter equation, it is useful to rewrite the Lax operators (\ref {lax})
in a coordinate representation.

Let us first recall the {\it position-momentum} Heisenberg algebra, generated by
elements $x_m$, $p_m$, $1\leq m \leq N$, satisfying:
\begin{eqnarray}
&&[x_m \, , \, x_n]=0\, , \nonumber \\
&&[ p_m \, ,\, p_n ]=0\, , \nonumber \\
&&[ x_m \, ,\,  p_n]= \frac {i\pi \beta ^2}{2}\delta _{m,n} \, . \nonumber
\end{eqnarray}
The key observation is that we can realize the quantum generators
$V_m^{\pm}$ for $1\leq m \leq N$ (\ref {Ccorr1}, \ref{Ccorr2}) by using {\it
position} and {\it momentum} $x_m$, $p_m$:
\be
V_m^{\pm}= \pm (x_{m+1}- x_m) - p_m \, ,
\label{vbos}
\ee
where the algebra element $x_{N+1}$ is identified with $x_1$ or, although
unnecessary for the following, we may think of $x_h$ and $p_h$ ($h\in {\Bbb
Z}$) as $N$-periodic objects in $h$. In any case, it is easy to verify that
elements (\ref{vbos}) satisfy commutation relations (\ref{vrel1}-\ref {vrel3}).

Now, we may use the usual coordinate representation $\hat x_m$, $\hat p_m$
for the elements $x_m$, $p_m$, respectively, \cite{STF} to obtain a coordinate
representation for $V^{\pm}_m$.

Let us indicate by ${\cal H}$ the {\it enlarged} vector space consisting of the
$L^2({\Bbb R})$ functions and of the  distributions. Let us consider the
$N$-tensor product of ${\cal H}$, ${\cal T}({\cal H})={\cal H}\otimes \ldots
\otimes {\cal H}$.  The representative operators for the positions $\hat x_m$,
$1\leq m \leq N$, act multiplicatively on the vectors of ${\cal T}({\cal H})$:
\begin{equation} (\hat x_m \psi )(x_1, \ldots, x_N)=x_m \psi (x_1, \ldots ,
x_N) \, , \label {repx}
\end{equation}
while the representative operators for the momentums $\hat p_m$ act as derivations:
\begin{equation}
(\hat p_m \psi )(x_1, \ldots, x_N)=-\frac {i\pi \beta ^2}{2}\frac {\partial}{\partial
  x_m} \psi (x_1, \ldots , x_N) \, .
\label {repp}
\end{equation}
Both representatives are well defined on the {\it enlarged} space ${\cal
T}({\cal H})$ and, for sake of simplicity, we have used the same
symbol for the algebra element $x_m$ and for the independent variable of the
$m$-th ${\cal H}$ space. Since in the following we will never write explicitly
abstract elements of the position-momentum Heisenberg algebra, this will cause
no confusion. In general, in order to have simple notations, from now on we
will indicate with the same symbol all the algebra elements and all their
representative operators, as the distinction will arise from the context. This
implies an accidental coincidence of the symbols for the independent position
variable and the corresponding position representative operator, but we will
never write explicitly position representative operators in the following:
$x_m$ will always indicate exclusively the position variable.

\medskip

From (\ref{vbos}) and from (\ref {repx}, \ref {repp}) we have the
following representation of $V_m^{\pm}$ ($1\leq m \leq N$)
\be
(V^{\pm}_m \psi )(x_1, \ldots, x_N)=\left [\pm (x_{m+1}-x_m)+\frac {i\pi \beta ^2}{2}
\frac {\partial}{\partial x_m}\right ]\psi (x_1, \ldots, x_N) \, , \nn
\ee
where the independent variable inherits the identification $x_{N+1}=x_1$ from
the algebra element. This implies that the operator representatives of
$W_m^{\pm}=e^{iV^{\pm}_m}$ ($1\leq m \leq N$) are defined as unitary operators
acting on ${\cal T}({\cal H})$ as follows:
\be
[W_m^{\pm}\psi](x_1, \ldots ,x_N)=e^{\pm i(x_{m+1}-x_m)}
e^{\pm \frac {i\pi \beta ^2}{4}}\psi (x_1, \ldots ,
x_m -\frac {\pi \beta ^2}{2}, \ldots , x_N) \, ,
\label {Wrep}
\ee
with the usual prescription $x_{N+1}=x_1$, for $m=N$.

Finally, inserting (\ref {Wrep}) in (\ref {lax}) we obtain a coordinate representation for
the left and right Lax operators. Since the entries of the Lax operators depend on $W_m^{\pm}$, they are well defined unitary operators acting on the whole
${\cal T}({\cal H})$.

Let us finally remark that the definition of the representation is a crucial problem in usual ABA and {\it a fortiori} in our non-ultralocal case: actually, this is the origin of the mistake in \cite {Kun}.

\section{Algebraic Bethe Ansatz in the conformal case.}
\setcounter{equation}{0}

\subsection{The left monodromy matrix.}
In this Subsection we will consider the left conformal monodromy matrix
(\ref{leftmon}) whose entries are defined by:
\begin{equation}
M(\lambda )=L_N(\lambda ) \ldots L_1(\lambda) \equiv
\left ( \begin{array}{cc}  A(\lambda )& B(\lambda)
\\ C(\lambda)  & D(\lambda) \\ \end{array} \right ) \, .
\end{equation}
We will consider the case of even number of sites, that is $N\in 2{\Bbb N}$, and we will write the Bethe equations, eigenvalues and eigenvectors of its
transfer matrix by developing an extension of ABA techniques. In fact, the
usual ABA grounds on the usual Yang-Baxter equation and hence we have to
modify it in such a way that we can use efficiently the braided equation. This can be rigorously done using the coordinate
representation given in the previous Section.

Let us define the fused Lax operator and its entries as: $k \in 2{\Bbb N}$,
$2\leq k \leq N$,
\begin{equation}
F_k(\lambda)\equiv L_k(\lambda )L_{k-1}(\lambda )\equiv \left ( \begin{array}{cc}
F_{11}^{(k,k-1)}(\lambda)&  F_{12}^{(k,k-1)}(\lambda)
\\F_{21}^{(k,k-1)}(\lambda)&  F_{22}^{(k,k-1)}(\lambda)\\ \end{array}\right )
\, ,
\label{fusl}
\end{equation}
and hence, from the definition (\ref{lax}), the entries are given by:
\begin{eqnarray}
 F_{11}^{(k,k-1)}(\lambda)&=& (W_k^-)^{-1}(W_{k-1}^-)^{-1}+\Delta ^2
\lambda ^2 W_k^+(W_{k-1}^+)^{-1} \, ,\label {fusent1}\\
F_{12}^{(k,k-1)}(\lambda)&=& \Delta \lambda [ (W_k^-)^{-1}W_{k-1}^+ +
W_k^+W_{k-1}^-] \, ,
\label {fusent2} \\
F_{21}^{(k,k-1)}(\lambda)&=& \Delta \lambda [ (W_k^+)^{-1}(W_{k-1}^-)^{-1}+
W_k^-(W_{k-1}^+)^{-1}] \, ,
\label {fusent3} \\
F_{22}^{(k,k-1)}(\lambda)&=& W_k^- W_{k-1}^- +\Delta ^2 \lambda ^2
(W_k^+)^{-1}W_{k-1}^+ \, .
\label {fusent4}
\end{eqnarray}

Let us go now to the coordinate representation (\ref{Wrep}). The fused Lax
operator entries (\ref{fusent1},  \ref{fusent3}, \ref{fusent4}) act as follows
on the representation space ${\cal T}({\cal H})$:
\begin{eqnarray}
&&[F_{11}^{(k,k-1)}(\lambda)\psi ](x_1, \ldots , x_N)=e^{i(x_{k+1}-x_{k-1})}\psi
(x_1, \ldots, x_{k-1}^+ , x_{k}^+, \ldots , x_N)+\nonumber \\
&&+\Delta ^2 \lambda ^2 q^{-1}
e^{i(x_{k+1}+x_{k-1}-2x_k)}\psi (x_1, \ldots, x_{k-1}^+ , x_{k}^-, \ldots , x_N)
\, , \label{coorep1} \\
\nn \\
&&[F_{22}^{(k,k-1)}(\lambda)\psi ](x_1, \ldots , x_N)=e^{-i(x_{k+1}-x_{k-1})}
\psi (x_1, \ldots, x_{k-1}^- , x_{k}^-, \ldots , x_N)+ \nonumber \\
&&+\Delta ^2 \lambda ^2 q^{-1}
e^{-i(x_{k+1}+x_{k-1}-2x_k)}\psi (x_1,\ldots, x_{k-1}^-, x_{k}^+, \ldots , x_N)
\, , \label {coorep2} \\
\nn \\
&&[F_{21}^{(k,k-1)}(\lambda)\psi ](x_1, \ldots , x_N)=\Delta \lambda
q^{-\frac{1}{2}} \left [ e^{-i(x_{k+1}-x_{k-1})}\psi (x_1, \ldots, x_{k-1}^+
, x_{k}^-, \ldots , x_N )+\right . \nonumber \\
&&\left . +e^{-i(x_{k+1}+x_{k-1}-2x_k)}\psi (x_1, \ldots, x_{k-1}^+,
x_{k}^+, \ldots , x_N) \right ] \, ,
\label {coorep3}
\end{eqnarray}
where for sake of conciseness we have defined
\begin{equation}
x^\pm_k \equiv x_k \pm \pi \beta ^2/2 \, ,
\end{equation}
and, of course, the variable $x_{N+1}$ is identified with $x_1$. Notice from
the previous formul\ae \, that the action of the operator $F_{ij}^{(k,k-1)}$ is
not confined on the coordinate $(x_k,x_{k-1})$ and is therefore called {\it
non-ultralocal}.

In order to carry on the usual ABA procedure, we have to find the
so-called {\it pseudovacuum} states.

\begin{definition}
In a fixed representation a pseudovacuum or false vacuum is a vector which is
simultaneous
eigenstate of the diagonal elements $A(\la)$ and $D(\la)$ of the monodromy
matrix and which is annihilated by the off-diagonal element $C(\la)$, for every $\la \in
{\Bbb C}$.
\end{definition}

We are now in the position to show that in the coordinate
representation space ${\cal T}({\cal H})$ the pseudovacua are given by
\begin{equation}
\Omega (x_1, \ldots , x_N)=
\prod _{{\stackrel{k=2}{k\in 2{\Bbb Z}}}}^N f(x_{k-1}-x_k)  \, ,
\label{pse}
\end{equation}
where $f$ is an element of ${\cal H}\otimes {\cal H}$, depending on the difference of
the coordinates and satisfying the shift property:
\begin{equation}
f(x+\pi \beta ^2)=-e^{-2ix}f(x)\, .
\label{thepro}
\end{equation}
The functional equation (\ref{thepro}) possesses in general infinite
solutions, for instance
\begin{equation}
f(x)={\mbox {exp}} \left ( -\frac {ix^2}{\pi \beta ^2}+ix+\frac {ix}{\beta ^2}
\right)
\label{funzione}
\end{equation}
and functions obtained from it by multiplication by a periodic function with
period $\pi \beta^2$. As we will show, however, every
solution of (\ref {thepro}) gives a pseudovacuum with the same eigenvalue for
$A$ and $D$. Hence, we do not need to single out any specific solution
of (\ref{thepro}).

\medskip

The proof of the fact that (\ref{pse}) with (\ref{thepro}) is a pseudovacuum,
relies on annihilation properties following immediately from (\ref{coorep3})
and (\ref{thepro}):
\be
[F_{21}^{(k,k-1)}(\lambda)\Omega ](x_1, \ldots , x_N)=0 \, .
\label{offdia}
\ee
Indeed, let us consider the expressions of $A$, $D$, $C$ in terms of the
elements of the fused Lax operator. For example, if $N=6$ we have
(understanding the  dependence on the spectral parameter):
\begin{eqnarray}
A&=&F_{11}^{(6,5)}\left [
F_{11}^{(4,3)}F_{11}^{(2,1)}+F_{12}^{(4,3)}F_{21}^{(2,1)}\right ] +
F_{12}^{(6,5)}\left [
F_{21}^{(4,3)}F_{11}^{(2,1)}+F_{22}^{(4,3)}F_{21}^{(2,1)}\right ], \nonumber
\\
D&=&F_{21}^{(6,5)}\left [
F_{11}^{(4,3)}F_{12}^{(2,1)}+F_{12}^{(4,3)}F_{22}^{(2,1)}\right ] +
F_{22}^{(6,5)}\left [
F_{21}^{(4,3)}F_{12}^{(2,1)}+F_{22}^{(4,3)}F_{22}^{(2,1)}\right ] ,\label{N6}
\\
C&=&F_{21}^{(6,5)}\left [ F_{11}^{(4,3)}F_{11}^{(2,1)}+
F_{12}^{(4,3)}F_{21}^{(2,1)}\right ] + F_{22}^{(6,5)}\left [
F_{21}^{(4,3)}F_{11}^{(2,1)}+F_{22}^{(4,3)} F_{21}^{(2,1)}\right ] \, .
\nonumber
\end{eqnarray}
Now, we prove, by using the $W$'s exchange algebra (\ref
{Wrel}), some very fundamental exchange relations between the
$F_{ij}^{(k,k-1)}(\la)$ ($k\in 2{\Bbb N}$, $2\leq k \leq N$) -- not
necessarily in a representation:

\begin{itemize}

\item exchange $(21)$-$(11)$
\ba
F_{21}^{(k+2,k+1)}(\la) F_{11}^{(k,k-1)}(\la^{\prime})&=&q^{-\frac {1}{2}}
 F_{11}^{(k,k-1)}(\la^{\prime}) F_{21}^{(k+2,k+1)}(\la)  \, , \nn \\
F_{21}^{(N,N-1)}(\la) F_{11}^{(2,1)}(\la^{\prime})&=&q^{-\frac {1}{2}}
 F_{11}^{(2,1)}(\la^{\prime}) F_{21}^{(N,N-1)}(\la) \, ;
\label{brafus1}
\ea

\item exchange $(21)$-$(12)$
\ba
F_{21}^{(k+2,k+1)}(\la) F_{12}^{(k,k-1)}(\la^{\prime})&=&q^{-\frac {1}{2}}
 F_{12}^{(k,k-1)}(\la^{\prime}) F_{21}^{(k+2,k+1)}(\la)  \, , \nn \\
F_{21}^{(N,N-1)}(\la) F_{12}^{(2,1)}(\la^{\prime})&=&q^{\frac {1}{2}}
 F_{12}^{(2,1)}(\la^{\prime}) F_{21}^{(N,N-1)}(\la) \, ;
\label{brafus2}
\ea

\item exchange $(21)$-$(22)$
\ba
F_{21}^{(k+2,k+1)}(\la) F_{22}^{(k,k-1)}(\la^{\prime})&=&q^{\frac {1}{2}}
 F_{22}^{(k,k-1)}(\la^{\prime}) F_{21}^{(k+2,k+1)}(\la)  \, , \nn \\
F_{21}^{(N,N-1)}(\la) F_{22}^{(2,1)}(\la^{\prime})&=&q^{\frac {1}{2}}
 F_{22}^{(2,1)}(\la^{\prime}) F_{21}^{(N,N-1)}(\la) \, ;
\label{brafus3}
\ea

\item commutation if ($k^\prime\in 2{\Bbb N}$, $2\leq k^\prime \leq
N$) $2<|k-k^\prime|<N-2$
\be [ F_{ij}^{(k,k-1)}(\la) \, , \, F_{i^\prime
j^\prime}^{(k^\prime , k^\prime -1)} (\la ^\prime)]=0   \, .
\label{brafus4}
\ee

\end{itemize}

Consequently, through the exchange properties (\ref{brafus1}-\ref{brafus4}), we
can bring all the factors $F_{21}^{(k,k-1)}$ to the right of the addenda in the
expressions of $A(\la)$, $D(\la)$, $C(\la)$. The following action of $A(\la)$,
$D(\la)$, $C(\la)$ on the state $\Omega$ (\ref{pse}) is a consequence of their
form (see as example formul\ae \, (\ref{N6}) in the case $N=6$) and of
annihilation properties (\ref{offdia}):
\begin{eqnarray}
A(\lambda )\Omega &=& \prod _{{\stackrel{k=2}{k\in 2{\Bbb Z}}}}^{{\stackrel
{N}{\leftarrow}}} F_{11}^{(k,k-1)} (\lambda)\Omega \, , \nn \\
D(\lambda )\Omega
&=& \prod _{{\stackrel{k=2}{k\in 2{\Bbb Z}}}}^{{\stackrel {N}{\leftarrow}}}
F_{22}^{(k,k-1)} (\lambda)\Omega \, , \nn \\
\nn \\
C(\lambda )\Omega &=& 0 \, ,
\label{Com}
\end{eqnarray}
where the arrow $\leftarrow$ indicates the verse of increasing indices in the
ordered product. We are left with proving that $\Omega$ is a simultaneous
eigenvector of $A(\la)$ and $D(\la)$: this will be realized by the following
theorem and its corollary.

\begin{theorem}
\label{conf1}
The action of the ordered product of diagonal elements of the fused
Lax operators (\ref{fusl}) on the states (\ref{pse}) is the following ($k\leq
N$)
\begin{eqnarray} &&\Bigl [\prod _{{\stackrel{h=2}{h\in 2{\Bbb
Z}}}}^{{\stackrel {k}{\leftarrow}}}  F_{11}^{(h,h-1)} (\lambda)\Omega
\Bigr](x_1,\ldots ,x_N)=e^{i(x_{k+1}-x_1)}q^{-\frac {1}{2}\left (\frac
{k}{2}-1\right)}(1-\Delta ^2 \lambda ^2 q^{-1})^{\frac {k}{2}} \Omega
(x_1,\ldots ,x_N) \nonumber \\ &&\Bigl [\prod _{{\stackrel{h=2}{h\in 2{\Bbb
Z}}}}^{{\stackrel {k} {\leftarrow}}}F_{22}^{(h,h-1)}(\lambda )\Omega \Bigr
](x_1,\ldots ,x_N)=e^{-i(x_{k+1}-x_1)}q^{-\frac {1}{2}\left (\frac
{k}{2}-1\right)}(1-\Delta ^2 \lambda ^2 q)^{\frac {k}{2}} \Omega (x_1,\ldots
,x_N) . \nonumber
\end{eqnarray}
\end{theorem}

{\bf Proof}: Let us show by induction the first formula. For $k=2$ it
follows from (\ref{coorep1}) and from the shift property (\ref{thepro}) .
For general $k\leq N$ we have from (\ref {coorep1}):
\begin{eqnarray}
&&\Bigl [\prod _{{\stackrel{h=2}{h\in 2{\Bbb Z}}}}^{{\stackrel {k}{\leftarrow}}}
 F_{11}^{(h,h-1)} (\lambda)\Omega \Bigr ](x_1, \ldots , x_N)= \nonumber \\
&& = e^{i(x_{k+1}-x_{k-1})}
\Bigl [\prod _{{\stackrel{h=2}{h\in 2{\Bbb Z}}}}^{{\stackrel {k-2}{\leftarrow}}}
 F_{11}^{(h,h-1)} (\lambda)\Omega \Bigr ](x_1, \ldots, x_{k-2},x_{k-1}^+,x_{k}^+,x_{k+1}, \ldots , x_N)+\nonumber \\
&&+\Delta ^2 \lambda ^2 q^{-1}
e^{i(x_{k+1}+x_{k-1}-2x_k)}\Bigl [\prod _{{\stackrel{h=2}{h\in 2{\Bbb Z}}}}^{{\stackrel {k-2}{\leftarrow}}}
 F_{11}^{(h,h-1)} (\lambda)\Omega \Bigr ](x_1, \ldots, x_{k-2},x_{k-1}^+,x_{k}^-, x_{k+1},\ldots , x_N ) \, . \nonumber
\end{eqnarray}
Applying the inductive hypothesis
we get:
\begin{eqnarray}
&&\Bigl [\prod _{{\stackrel{h=2}{h\in 2{\Bbb Z}}}}^{{\stackrel {k}{\leftarrow}}}
 F_{11}^{(h,h-1)} (\lambda)\Omega \Bigr ](x_1, \ldots , x_N)= \nonumber \\
&& = e^{i(x_{k+1}-x_{k-1})}
e^{i(x_{k-1}^+-x_1)}q^{-\frac {1}{2}\left (\frac {k}{2}-2\right)}(1-\Delta ^2 \lambda ^2 q^{-1})^{\frac {k}{2}-1} \Omega (x_1,\ldots ,x_N) +\nonumber \\
&& +\Delta ^2 \lambda ^2 q^{-1} e^{i(x_{k+1}+x_{k-1}-2x_k)}
e^{i(x_{k-1}^+-x_1)}q^{-\frac {1}{2}\left (\frac {k}{2}-2\right)}(1-\Delta ^2 \lambda ^2 q^{-1})^{\frac {k}{2}-1} \, \cdot \nonumber \\
&& \cdot \, \Omega (x_1, \ldots, x_{k-2}, x_{k-1}^+, x_k^-, x_{k+1}, \ldots ,x_N) \, .
\end{eqnarray}
The use of the shift property (\ref {thepro}) in the last term gives:
\begin{equation}
\Omega (x_1, \ldots, x_{k-2}, x_{k-1}^+, x_k^-, x_{k+1}, \ldots ,x_N)=
-e^{2i(x_k-x_{k-1})}\Omega (x_1, \ldots , x_N) \, .
\end{equation}
Hence the two terms of the right hand side are
proportional. After gathering them, we get the first formula of Theorem \ref
{conf1}.

The second formula follows in an analogous way, after using the shift property
(\ref{thepro}) in the form:
\begin{equation}
f(x-\pi\beta ^2)=-e^{2i(x-\pi \beta ^2)}f(x) \, .  \nonumber
\end{equation}
\finproof

\medskip

\begin{corollary} \label{coroconf1} The states (\ref {pse}) are eigenvectors
of the elements $A(\lambda)$ and $D(\lambda)$ of the left conformal monodromy
matrix (\ref {leftmon}). The corresponding common eigenvalues are given by the
following formul\ae:  \begin{eqnarray}  &&[A(\lambda )\Omega ]=q^{-\frac
{1}{2}\left (\frac {N}{2}-1\right)} (1-\Delta ^2 \lambda ^2 q^{-1})^{\frac
{N}{2}} \Omega  \equiv \rho _N (\lambda ) \Omega  \, ,  \label {Aom} \\
&&[D(\lambda )\Omega ]=q^{-\frac {1}{2}\left (\frac {N}{2}-1\right)}(1-\Delta
^2 \lambda ^2 q)^{\frac {N}{2}} \Omega  \equiv \sigma _N (\lambda ) \Omega \,
. \label {Dom} \end{eqnarray} \end{corollary} {\bf Proof}: The proof follows
from Theorem \ref {conf1} for $k=N$, remembering  that $x_{N+1}=x_1$.
\finproof

\medskip

\noindent
Eventually, formul\ae \, (\ref{Com}), (\ref{Aom}) and (\ref{Dom}) show that
the states (\ref{pse}) are pseudovacua of the monodromy matrix
(\ref{leftmon}) with the same $A(\lambda)$ and $D(\lambda)$ eigenvalues for
any $f(x)$ verifying (\ref{thepro}). Nevertheless, we need to notice that the two site state
$f(x_{k-1}-x_k)$ is not a pseudovacuum for $A^{(k,k-1)}\equiv
F_{11}^{(k,k-1)}$, $D^{(k,k-1)}\equiv F_{22}^{(k,k-1)}$, $C^{(k,k-1)}\equiv
F_{21}^{(k,k-1)}$: this property is quite rare and called {\it non-ultralocality
of the pseudovacuum}.

\medskip

Let us derive now the Bethe Ansatz equations. From (\ref{pmon2}) it follows
that the left conformal monodromy matrix (\ref{leftmon}) satisfies the
braided Yang-Baxter relation:
\begin{equation}
R_{ab}\left (\frac {\lambda}{\lambda ^\prime} \right)M _a(\lambda
)Z_{ab}^{-1}M _b(\lambda ^\prime)=M_b(\lambda ^\prime)Z_{ab}^{-1}M_a
(\lambda )R_{ab}\left (\frac {\lambda }{\lambda ^\prime}\right ) \, .
\label {rll}
\end{equation}
which contains implicitly these exchange rules between $B(\la ^\prime)$ and $A(\la)$, $D(\la)$ respectively:
\begin{eqnarray}
A(\lambda)B(\lambda ^\prime)&=&\frac {q^{-1}}{a\left(\frac
  {\lambda ^\prime}{\lambda}\right)}B(\lambda ^\prime)A(\lambda) -
{q^{-1}}\frac {b\left(\frac
  {\lambda ^\prime}{\lambda}\right)}{a\left(\frac
  {\lambda ^\prime}{\lambda}\right)}B(\lambda)A(\lambda ^\prime) \, ,
\label{exch1} \\
D(\lambda)B(\lambda ^\prime)&=&\frac {q}{a\left(\frac
  {\lambda}{\lambda ^\prime}\right)}B(\lambda ^\prime)D(\lambda) -
{q}\frac {b\left(\frac
  {\lambda}{\lambda ^\prime}\right)}{a\left(\frac
  {\lambda}{\lambda ^\prime}\right)}B(\lambda)D(\lambda ^\prime) \, .
\label{exch2}
\end{eqnarray}
In equations (\ref{exch1}, \ref{exch2}) we have defined for sake of
conciseness:
\begin{equation}
a(\xi )=\frac {\xi ^{-1}-\xi }{q^{-1}\xi ^{-1}-q\xi } \, , \quad b(\xi)=\frac
{q^{-1}-q}{q^{-1}\xi ^{-1}-q\xi } \, .
\label{ab}
\end{equation}
Note the presence of the factors $q^{\pm 1}$ in expressions (\ref{exch1},
\ref{exch2}): they come from the matrix $Z_{ab}$ and represent the
contribution to exchange relations coming from non-ultralocality. Now, as
usual we build Bethe states
\begin{equation}
\Psi (\lambda _1, \ldots, \lambda _l)=\prod _{r=1}^lB(\lambda
_r)\Omega
\,
\label{psi}
\end{equation}
acting on a pseudovacuum with the {\it creators of pseudoparticles}
$B(\la_r)$, without care about ordering because of the commuting property
encoded in the braided Yang-Baxter equation:
\be
[ B(\la), B(\la^{\prime}) ]=0.
\ee
From (\ref {exch1}, \ref {exch2}) we find the action of
$A(\lambda)$ and $D(\lambda)$ on Bethe states:
\begin{eqnarray}
A(\lambda)\Psi (\lambda _1, \ldots, \lambda _l)&=& q^{-l}\prod
_{r=1}^l\frac {1}{a(\frac {\lambda_r}{\lambda})} \rho _N(\lambda)
\Psi (\lambda _1, \ldots, \lambda _l) + \ldots \label {A}\\
D(\lambda)\Psi (\lambda _1, \ldots, \lambda _l)&=& q^{l}\prod
_{r=1}^l\frac {1}{a(\frac {\lambda}{\lambda_r})} \sigma _N(\lambda )
\Psi (\lambda _1, \ldots, \lambda _l) + \ldots \, .
\label{D}
\end{eqnarray}
The dots in (\ref {A}, \ref {D}) indicate extra terms which are not proportional
to the state (\ref{psi}). Hence, in general, states (\ref {psi}) are not
eigenstates of the $\lambda$-dependent transfer matrices
$T(\lambda)=A(\lambda)+D(\lambda)$. This is true if and only if the set of
complex numbers $\{ \lambda _1, \ldots , \lambda _l \}$ satisfy the following
Bethe Equations (BE's):
\begin{equation} q^{-l}\prod _{{\stackrel{r=1}{r\not=s}}}
^l\frac {1}{a\left(\frac {\lambda _r}{\lambda _s}
\right)} \rho _N(\lambda_s)=
q^{l}\prod _{{\stackrel{r=1}{r\not=s}}}^l\frac {1}{a\left(\frac {\lambda _s}
{\lambda _r}
\right)} \sigma _N(\lambda_s )
\, .
\label{bethel1}
\end{equation}
By using the expressions for $ \rho _N(\lambda)$ and $\sigma _N (\lambda )$
coming from (\ref{Aom}, \ref{Dom}) and for $a(\la)$ coming
from (\ref{ab}), we can rewrite the BE's as follows:
\begin{equation}
q^{-2l}\prod _{{\stackrel{r=1}{r\not=s}}}^l \frac {q\lambda _r^2-q^{-1}\lambda
  _s^2}{q^{-1}\lambda _r^2-q\lambda _s^2}=\left (
\frac {1-\Delta ^2 \lambda _s ^2 q}{1-\Delta ^2\lambda _s ^2q^{-1}}
\right)^{N/2}
\, .
\label{bethel2}
\end{equation}
The definition
\begin{equation}
\Delta \lambda _r \equiv e^{\alpha _r} \,
\end{equation}
allows us to rewrite BE's (\ref{bethel2}) in the more diffuse trigonometric
form:
\begin{equation}
\prod _{{\stackrel{r=1}{r\not=s}}}^l\frac {\sinh (\alpha _s -\alpha _r +i\pi
\beta ^2)}{\sinh (\alpha _s -\alpha _r -i\pi \beta ^2)} = \left [ \frac
{\sinh \left (\alpha _s -\frac {i\pi \beta ^2}{2}\right)} {\sinh \left (\alpha
_s +\frac {i\pi \beta ^2}{2}\right)}\right ]^{N/2} e^{-\frac {i\pi \beta
^2}{2} N-2i\pi \beta ^2 l} \, .
\label{bethel3}
\end{equation}

Eventually, let us deduce, from equations (\ref{A},\ref{D}), the eigenvalues
of the left transfer matrix $T(\lambda)\equiv {\mbox {Tr}}\, M(\lambda )$,
relatively to Bethe states (\ref{psi}), (\ref{bethel2}):
\begin{equation}  \Lambda (\lambda, \{\lambda _r\})=q^{-l}\prod _{r=1}^l\frac
{1}{a(\frac {\lambda_r}{\lambda})} \rho _N(\lambda)+ q^{l}\prod _{r=1}^l\frac
{1} {a(\frac {\lambda}{\lambda_r})}\sigma _N(\lambda )\, .
\label{eigen1}
\end{equation}
By using the expressions for $\rho_N(\lambda)$ and $\sigma_N(\lambda )$
coming from (\ref{Aom}, \ref{Dom}), we write (\ref{eigen1}) in the following way:
\begin{eqnarray}
\Lambda (\lambda , \{ \lambda _r\})&=&
q^{-l}\prod _{r=1}^l
\frac {q^{-1}\lambda
  ^2-q\lambda _r^2}{\lambda ^2-\lambda
  _r^2}q^{-\frac {1}{2}\left (\frac {N}{2}-1\right)}(1-\Delta ^2\lambda ^2q^{-1})^{N/2}+\nonumber \\
 &+&q^{l}\prod _{r=1}^l
\frac {q\lambda ^2-q^{-1}\lambda _r^2}{\lambda ^2-\lambda
  _r^2}q^{-\frac {1}{2}\left (\frac {N}{2}-1\right)}(1-\Delta ^2\lambda ^2q)^{N/2}
\, .
\label{T}
\end{eqnarray}

Let us produce some comments about the results of this subsection. The BE's
(\ref{bethel3}) are the equations for a spin chain of spin $-\frac{1}{2}$
with, in addition, the {\it twist} $e^{-\frac {i\pi \beta ^2}{2} N-2i\pi \beta
^2 l}$. Instead, in paper \cite{Kun} they turn out to be of different signs
(spin $+\frac{1}{2}$ chain), because of an inconsistent definition of the
pseudovacuum, affecting also the final expressions of the eigenvalues. As far
as we know, the presence of the $l$-dependent twist appearing in the BE's is a
new feature and is a direct consequence of non-ultralocality, encoded in the
$Z_{ab}$ matrix. In view of the fact that this twist depends on the number of
the Bethe roots (the solutions of the BE's), it will be said as {\it
dynamically generated}. The form of the eigenvalues of the transfer matrix
(\ref{T}) are as well those of a dynamically twisted $-\frac{1}{2}$ spin
chain. A similarly generated twist appeared in \cite{FTi} in the case of a CFT
-- Liouville theory -- but it is only depending on the number of sites $N$.
Besides, in paper \cite{FTi} a detailed analysis has been carried out to
conjecture a one-to-one correspondence between Bethe states and squares in Kac
table of minimal CFT's.  These facts lead us to think that the (cylinder)
continuum limit of the equations (\ref{bethel3}), (\ref{T}) describes the
chiral sector of CFT's and their chiral IMI encoded in the transfer matrix. In
a forthcoming paper \cite{35} we will examine the (cylinder) continuum limit
for special values of $\beta^2$ corresponding to the very interesting case of
minimal CFT's in order prove this conjecture.

\subsection{Right monodromy matrix.}
We may repeat all the steps and considerations of the last subsection in the
case of the right conformal monodromy matrix (\ref{rightmon})
\begin{equation}
\bar M(\lambda )=\bar L_N(\lambda ^{-1}) \ldots \bar L_1(\lambda ^{-1}) \equiv
\left ( \begin{array}{cc}  \bar A(\lambda )& \bar B(\lambda)
\\ \bar C(\lambda)  & \bar D(\lambda) \\ \end{array} \right ) \, , \, \, N\in 2{\Bbb N}
\nonumber
\end{equation}
and hence we will briefly illustrate them.

Now, the fused Lax operator and its entries are defined by ($k \in 2{{\Bbb N}}
 , \, \, 2\leq k \leq N $)
\begin{equation}
\bar F_k(\lambda ^{-1})\equiv \bar L_k(\lambda ^{-1})\bar L_{k-1}
(\lambda ^{-1})\equiv\left ( \begin{array}{cc}  \bar F_{11}^{(k,k-1)}(\lambda
^{-1})&  \bar F_{12}^{(k,k-1)}(\lambda ^{-1}) \\\bar F_{21}^{(k,k-1)}(\lambda
^{-1})& \bar F_{22}^{(k,k-1)}(\lambda ^{-1})\\ \end{array}\right )
\label{rfusl}
\end{equation}
and hence from (\ref {lax}) the entries are explicitly:
\begin{eqnarray}
\bar F_{11}^{(k,k-1)}(\lambda ^{-1})&=&
(W_k^+)^{-1}(W_{k-1}^+)^{-1}+\Delta ^2 \lambda ^{-2} W_k^-(W_{k-1}^-)^{-1} \,
,\label {rfusent1}\\ \bar F_{12}^{(k,k-1)}(\lambda ^{-1})&=& \Delta
\lambda ^{-1}[ (W_k^+)^{-1}W_{k-1}^- + W_k^-W_{k-1}^+] \, , \label {rfusent2}
\\ \bar F_{21}^{(k,k-1)}(\lambda ^{-1})&=& \Delta \lambda ^{-1}[
(W_k^-)^{-1}(W_{k-1}^+)^{-1}+ W_k^+(W_{k-1}^-)^{-1}] \, ,\label {rfusent3} \\
\bar F_{22}^{(k,k-1)}(\lambda ^{-1})&=& W_k^+ W_{k-1}^+ +\Delta ^2
\lambda^{-2} (W_k^-)^{-1}W_{k-1}^- \, . \label {rfusent4} \end{eqnarray}

In the coordinate representation (\ref{Wrep}) the fused Lax operator entries
(\ref{rfusent1}, \ref{rfusent3}, \ref{rfusent4}) act on the space ${\cal
T}({\cal H})$ as follows:
\begin{eqnarray}
&&[\bar F_{11}^{(k,k-1)}(\lambda ^{-1})\psi ](x_1, \ldots , x_N)=e^{-i(x_{k+1}-
x_{k-1})}\psi (x_1, \ldots, x_{k-1}^+,x_{k}^+, \ldots , x_N)+\nonumber \\
&&+\Delta ^2 \lambda ^{-2} q \, e^{-i(x_{k+1}+x_{k-1}-2x_k)}\psi (x_1, \ldots,
x_{k-1}^+,x_{k}^-, \ldots , x_N)
\, , \label{rcoorep1} \\
\nn \\
&&[\bar F_{22}^{(k,k-1)}(\lambda ^{-1})\psi ](x_1, \ldots , x_N)=e^{i(x_{k+1}-
x_{k-1})}\psi (x_1, \ldots, x_{k-1}^-,x_{k}^-, \ldots , x_N)+ \nonumber \\
&&+\Delta ^2 \lambda ^{-2} q \,
e^{i(x_{k+1}+x_{k-1}-2x_k)}\psi (x_1, \ldots, x_{k-1}^-,x_{k}^+, \ldots , x_N)
\, , \label{rcoorep2} \\
\nn \\
&&[\bar F_{21}^{(k,k-1)}(\lambda ^{-1})\psi ](x_1, \ldots , x_N)=\Delta \lambda
^{-1} q^{\frac {1}{2}} \left [ e^{i(x_{k+1}-x_{k-1})}\psi (x_1, \ldots,
x_{k-1}^+, x_{k}^-,\ldots ,x_N )+\right . \nonumber \\
&&\left . +e^{i(x_{k+1}+x_{k-1}-2x_k)}\psi (x_1, \ldots, x_{k-1}^+,  x_{k}^+,
\ldots , x_N) \right ] \, ,
\label{rcoorep3}
\end{eqnarray}
where again $x_{N+1}$ is to be meant as $x_1$.

We now show that in the coordinate representation space the pseudovacua are
given by:
\begin{equation}
\bar \Omega (x_1, \ldots , x_N)=\prod_{{\stackrel{k=2}{k\in 2{\Bbb Z}}}}^N
\bar f( x_{k-1}-x_k)\, ,
\label {rpse}
\end{equation}
where $\bar f(x)$ is characterized by the shift property:
\begin{equation}
\bar f(x+\pi \beta ^2)=-e^{2ix}\bar f(x) \, .
\label {rthepro}
\end{equation}
Any non-zero solution of (\ref{rthepro}) is given by a non-zero solution
of (\ref{thepro}) by inversion
\begin{equation}
\bar f(x)=f(x)^{-1} \, ,
\label{rfunzione}
\end{equation}
and the reverse. Hence, a particular solution is furnished by (\ref{funzione})
via (\ref{rfunzione}).

Again, from (\ref {rcoorep3}) and (\ref {rthepro}) it is easy to see the
basic annihilation properties:
\begin{equation}
[\bar F_{21}^{(k,k-1)}(\lambda ^{-1})\bar \Omega ](x_1, \ldots , x_N)=0 \, .
\label{roffdia}
\end{equation}
Then, we prove, as before, by using the $W$'s exchange algebra (\ref
{Wrel}), some very fundamental exchange relations between the
$\bar F_{ij}^{(k,k-1)}(\la)$ ($k\in 2{\Bbb N}$, $2\leq k \leq N$) -- not
necessarily in a representation):

\begin{itemize}

\item exchange $(21)$-$(11)$
\ba
\bar F_{21}^{(k+2,k+1)}(\la)\bar F_{11}^{(k,k-1)}(\la^{\prime})&=&q^{\frac
{1}{2}} \bar F_{11}^{(k,k-1)}(\la^{\prime}) \bar F_{21}^{(k+2,k+1)}(\la)  \,
 , \nn \\
\bar F_{21}^{(N,N-1)}(\la)\bar F_{11}^{(2,1)}(\la^{\prime})&=&q^{\frac {1}{2}}
\bar F_{11}^{(2,1)}(\la^{\prime}) \bar F_{21}^{(N,N-1)}(\la) \, ;
\label{rbrafus1}
\ea

\item exchange $(21)$-$(12)$
\ba
\bar F_{21}^{(k+2,k+1)}(\la) \bar F_{12}^{(k,k-1)}(\la^{\prime})&=&q^{\frac{1}{2}}
\bar F_{12}^{(k,k-1)}(\la^{\prime}) \bar F_{21}^{(k+2,k+1)}(\la) \, ,\nn \\
\bar F_{21}^{(N,N-1)}(\la)\bar F_{12}^{(2,1)}(\la^{\prime})&=&q^{-\frac {1}{2}}
\bar F_{12}^{(2,1)}(\la^{\prime}) \bar F_{21}^{(N,N-1)}(\la) \, ;
\label{rbrafus2}
\ea

\item exchange $(21)$-$(22)$
\ba
\bar F_{21}^{(k+2,k+1)}(\la)\bar F_{22}^{(k,k-1)}(\la^{\prime})&=&q^{-\frac {1}{2}}
\bar F_{22}^{(k,k-1)}(\la^{\prime}) \bar F_{21}^{(k+2,k+1)}(\la) \,  , \nn \\
\bar F_{21}^{(N,N-1)}(\la) \bar F_{22}^{(2,1)}(\la^{\prime})&=&q^{-\frac {1}{2}}
\bar F_{22}^{(2,1)}(\la^{\prime})\bar F_{21}^{(N,N-1)}(\la) \, ;
\label{rbrafus3}
\ea

\item commutation if ($k^\prime\in 2{\Bbb N}$, $2\leq k^\prime \leq
N$) $2<|k-k^\prime|<N-2$
\be
[\bar F_{ij}^{(k,k-1)}(\la) \, , \, \bar F_{i^\prime j^\prime}^{(k^\prime ,
k^\prime -1)} (\la ^\prime)]=0  \, .
\label{rbrafus4}
\ee

\end{itemize}

Consequently, we can bring all the  factors $\bar F_{21}^{(k,k-1)}(\la)$ to the
right of the addenda in the expressions of $\bar A(\la)$, $\bar D(\la)$, $\bar
C(\la)$. This terms annihilate (\ref{rpse}) and therefore
the action of $\bar A(\la)$, $\bar D(\la)$, $\bar C(\la)$ is reduced to:
\begin{eqnarray}
\bar A(\lambda )\bar \Omega
&=& \prod _{{\stackrel{k=2}{k\in 2{\Bbb Z}}}}^{\stackrel {N}{\leftarrow}} \bar
F_{11}^{(k,k-1)}(\lambda ^{-1}) \bar \Omega \, , \\ \bar D(\lambda )\bar
\Omega &=& \prod _{{\stackrel{k=2}{k\in 2{\Bbb Z}}}}^{\stackrel
{N}{\leftarrow}} \bar F_{22}^{(k,k-1)} (\lambda ^{-1})\bar \Omega \, ,\\
\bar C(\lambda )\bar \Omega &=& 0 \, .
\label {rCom}
\end{eqnarray}
Now, we can prove that (\ref{rpse}) are simultaneous eigenvectors of $\bar
A(\la)$ and $\bar D(\la)$.

\begin{theorem}
\label{conf2}
The action of the ordered product of the diagonal elements of the
fused Lax operators (\ref {rfusl}) on the states (\ref {rpse}) is ($k\leq N$):
\begin{eqnarray}
\Bigl [\prod _{{\stackrel{h=2}{h\in 2{\Bbb Z}}}}^{{\stackrel {k}{\leftarrow}}} \bar F_{11}^{(h,h-1)}(\lambda ^{-1})\bar \Omega \Bigr ](x_1, \ldots ,x_N)&=&e^{-i(x_{k+1}-x_1)}q^{\frac {1}{2}\left (\frac {k}{2}-1\right)}(1-\Delta ^2 \lambda ^{-2} q)^{\frac {k}{2}}
 \bar \Omega (x_1, \ldots ,x_N) \nonumber \\
\Bigl [\prod _{{\stackrel{h=2}{h\in 2{\Bbb Z}}}}^{{\stackrel {k}{\leftarrow}}} \bar F_{22}^{(h,h-1)}(\lambda ^{-1})\bar \Omega \Bigr ](x_1, \ldots ,x_N)&=&e^{i(x_{k+1}-x_1)}q^{\frac {1}{2}\left (\frac {k}{2}-1\right)}(1-\Delta ^2 \lambda ^{-2} q^{-1})^{\frac {k
}{2}} \bar \Omega (x_1, \ldots ,x_N)  \nonumber  \end{eqnarray}
\end{theorem}
{\bf Proof}: The proof is completely analogous to that of Theorem \ref{conf1}
and uses only the shift property (\ref{rthepro}).
\finproof

\medskip

\begin{corollary}
\label{coroconf2}
The states (\ref {rpse}) are eigenvectors of the elements $\bar A(\lambda)$ and
$\bar D(\lambda)$ of the right conformal monodromy matrix (\ref {rightmon}).
The corresponding common eigenvalues are given by the following formul\ae:
\begin{eqnarray}
&&[\bar A(\lambda )\bar \Omega ]=q^{\frac {1}{2}\left (\frac
{N}{2}-1\right)}(1-\Delta ^2 \lambda ^{-2} q)^{\frac {N}{2}} \bar \Omega   \,
,  \label {rAom} \\ &&[\bar D(\lambda )\bar \Omega ]=q^{\frac {1}{2}\left
(\frac {N}{2}-1\right)}(1-\Delta ^2 \lambda ^{-2} q^{-1})^{\frac {N}{2}} \bar
\Omega   \, .
\label {rDom}
\end{eqnarray}
\end{corollary}
{\bf Proof}: The proof follows from Theorem \ref{conf2} for $k=N$,
remembering that $x_{N+1}=x_1$.  \finproof

\medskip

Eventually, formul\ae \, (\ref{rCom}), (\ref{rAom}) and (\ref{rDom}) show that
the states (\ref{rpse}) are pseudovacua of the monodromy matrix
(\ref{rightmon}) with the same $\bar A(\lambda)$ and
$\bar D(\lambda)$ eigenvalues for any $\bar f(x)$ verifying
(\ref{rthepro}).

\medskip

Le us derive now the Bethe Ansatz equations. From (\ref{pmon2}) it follows that
the right conformal monodromy matrix (\ref{rightmon}) satisfies the braided
Yang-Baxter relation (\ref{rll}) with $Z_{ab}$ replaced by $Z_{ab}^{-1}$. This
relation contains these exchange rules between $\bar B(\la ^\prime)$ and $\bar
A(\la)$, $\bar D(\la)$ respectively:
\begin{eqnarray} \bar A(\lambda)\bar B(\lambda
^\prime)&=&\frac {q}{a\left(\frac   {\lambda  ^\prime  }{\lambda }\right)}\bar
B(\lambda ^\prime)\bar A(\lambda) - {q}\frac {b\left(\frac
  {\lambda ^\prime  }{\lambda }\right)}{a\left(\frac
  {\lambda  ^\prime }{\lambda }\right)}\bar B(\lambda)\bar A(\lambda ^\prime) \, ,
\label {rexch1} \\
\bar D(\lambda)\bar B(\lambda ^\prime)&=&\frac {q^{-1}}{a\left(\frac
  {\lambda }{\lambda  ^\prime }\right)}\bar B(\lambda ^\prime)\bar D(\lambda) -
{q^{-1}}\frac {b\left(\frac
  {\lambda }{\lambda  ^\prime }\right)}{a\left(\frac
  {\lambda }{\lambda  ^\prime  }\right)}\bar B(\lambda)\bar D(\lambda ^\prime) \, .
\label{rexch2}
\end{eqnarray}
We define the Bethe states in the usual way:
\begin{equation}
\bar \Psi (\lambda _1, \ldots, \lambda _l)=\prod _{r=1}^l\bar B(\lambda
_r)\bar \Omega
\, .
\label{rpsi}
\end{equation}
As a consequence of (\ref{rAom}, \ref{rDom}, \ref{rexch1}, \ref{rexch2}) the
Bethe Equations ({BE's}) for the right conformal monodromy matrix read as
\begin{equation}
q^{2l}\prod _{{\stackrel{r=1}{r\not= s}}}^l \frac
{q\lambda _r^2-q^{-1}\lambda   _s^2}{q^{-1}\lambda _r^2-q\lambda _s^2}=\left (
\frac {1-\Delta ^2 \lambda _s ^{-2} q^{-1}}{1-\Delta ^2\lambda _s ^{-2}q}
\right)^{N/2} \, ,
\label{rbethel2}
\end{equation}
or in a trigonometric form ($\Delta ^{-1}\lambda _r \equiv e^{\bar \alpha
_r}$): \begin{equation}
 \prod _{{\stackrel{r=1}{r\not=s}}}^l\frac {\sinh (\bar \alpha _s -\bar \alpha _r +i\pi \beta ^2)}{\sinh (\bar \alpha _s -\bar \alpha _r -i\pi \beta ^2)} = \left [ \frac
{\sinh \left (\bar \alpha _s -\frac {i\pi \beta ^2}{2}\right)} {\sinh \left (\bar \alpha _s +\frac {i\pi \beta ^2}{2}\right)}\right ]^{N/2} e^{\frac {i\pi \beta ^2}{2} N+2i\pi \beta ^2 l} \, .
\label{rbethel3}
\end{equation}
In addition the eigenvalues of the transfer matrix $\bar T(\lambda)\equiv
{\mbox {Tr}}\, \bar M(\lambda )$ are:
\begin{eqnarray}
\bar \Lambda (\lambda
, \{ \lambda _r\})&=& q^{l}\prod _{r=1}^l  \frac {q^{-1}\lambda^2-q\lambda _r^2}
{\lambda^2-\lambda_r^2}q^{\frac {1}{2}\left (\frac {N}{2}-1\right)}(1-\Delta^2
\lambda^{-2}q)^{N/2}+ \nonumber \\
&+&q^{-l}\prod _{r=1}^l \frac {q\lambda^2-q^{-1}\lambda_r^2}{\lambda ^2-
\lambda_r^2}q^{\frac{1}{2}\left( \frac{N}{2} - 1\right) }
(1-\Delta ^2\lambda ^{-2}q^{-1})^{N/2} \, .
\label{rT}
\end{eqnarray}

\medskip

We can comment the results of this subsection in an analogous way as we have
done at the end of the previous subsection, after taking into account the
change of left (chiral) into right (anti-chiral).

\section {Algebraic Bethe Ansatz in the off-critical case.}
\setcounter{equation}{0}
The minimal CFT's perturbed by the primary field $\Phi _{1,3}$ possess local
IMI, which are suitable {\it deformations} of those in left and right quantum
KdV theory in the continuum limit {\cite{SY,AZ,BLZ,FRS}. For this reason we
{\it couple together} left and right theories with the aim of describing a
lattice discretization (or better regularization) of perturbed CFT's.
Preserving integrability, we would like to conjecture that this coupling of
left (chiral) and right (anti-chiral) sectors is realized equivalently by the
monodromy matrices (\ref{Rightmon}) or (\ref {Leftmon}), which contain both
$L_m$ and $\bar L _m$ and verify braided Yang-Baxter relation. In this Section
we will diagonalize the associated transfer matrices by means of an extended
version of Algebraic Bethe Ansatz techniques.

Let us start with the monodromy matrix (\ref{Rightmon}) and let us defines its
entries ($N\in 4{{\Bbb N}}$)
\begin{equation}
{\bf M}(\lambda )=\prod _{i=1}^{\stackrel {N/4}{\leftarrow}}\bar F_{4i}\left
(\frac{\mu^{\frac {1}{2}}}{\lambda}\right) F_{4i-2}(\lambda \mu ^{\frac
  {1}{2}})\equiv \left ( \begin{array}{cc}  {\bf A}(\lambda;\mu)& {\bf
B}(\lambda;\mu) \\ {\bf C}(\lambda;\mu)  & {\bf D} (\lambda;\mu) \\ \end{array}
\right )  \, .
\label{MON}
\end{equation}
We want to write the eigenvectors and eigenvalues of the transfer matrix in terms of the
solutions (roots) of the Bethe Equations (BE's).

Remember that the fused Lax operators in (\ref {MON}) are
\begin{equation}
\bar F_{4i}(\lambda )= \bar L_{4i}(\lambda) \bar L_{4i-1}(
\lambda) \quad , \quad
F_{4i-2}(\lambda)=L_{4i-2}(\lambda)L_{4i-3}(\lambda ) \,
\end{equation}
and that their entries, defined by (\ref {rfusl}) for $\bar F_k$ and (\ref {fusl}) for $F_k$, are explicitly given by:
\begin{eqnarray}
\bar F_{11}^{(4i,4i-1)}(\lambda )&=& (W_{4i}^+)^{-1}(W_{4i-1}^+)^{-1}+\Delta ^2 \lambda ^2 W_{4i}^-(W_{4i-1}^-)^{-1} \, , \label {Fusent1}\\
\bar F_{12}^{(4i,4i-1)}(\lambda )&=& \Delta \lambda [ (W_{4i}^+)^{-1}W_{4i-1}^- +
W_{4i}^-W_{4i-1}^+] \, , \label {Fusent2} \\
\bar F_{21}^{(4i,4i-1)}(\lambda )&=& \Delta \lambda [ (W_{4i}^-)^{-1}(W_{4i-1}^+)^{-1}+ W_{4i}^+(W_{4i-1}^-)^{-1}]\, , \label {Fusent3} \\
\bar F_{22}^{(4i,4i-1)}(\lambda )&=& W_{4i}^+ W_{4i-1}^+ +\Delta ^2 \lambda ^2 (W_{4i}^-)^{-1}W_{4i-1}^- \, , \label {Fusent4} \\
F_{11}^{(4i-2,4i-3)}(\lambda )&=& (W_{4i-2}^-)^{-1}(W_{4i-3}^-)^{-1}+\Delta ^2 \lambda ^2 W_{4i-2}^+(W_{4i-3}^+)^{-1} \, , \label {Fusent5}\\
F_{12}^{(4i-2,4i-3)}(\lambda )&=& \Delta \lambda [ (W_{4i-2}^-)^{-1}W_{4i-3}^+ + W_{4i-2}^+W_{4i-3}^-] \, , \label {Fusent6} \\
F_{21}^{(4i-2,4i-3)}(\lambda )&=& \Delta \lambda [ (W_{4i-2}^+)^{-1}(W_{4i-3}^-)^{-1}+ W_{4i-2}^-(W_{4i-3}^+)^{-1}]\, , \label {Fusent7} \\
F_{22}^{(4i-2,4i-3)}(\lambda )&=& W_{4i-2}^- W_{4i-3}^- +\Delta ^2 \lambda ^2 (W_{4i-2}^+)^{-1}W_{4i-3}^+ \, . \label {Fusent8}
\end{eqnarray}

We go now to the coordinate representation. Actually, we have already written
how the operators representatives of (\ref{Fusent1}, \ref {Fusent3}, \ref {Fusent4}) and of (\ref{Fusent5}, \ref {Fusent7}, \ref {Fusent8}) act on the
coordinate space ${\cal T}({\cal H})$ in formul\ae \, (\ref{coorep1}-\ref{coorep3}) and
(\ref{rcoorep1}-\ref{rcoorep3}). These are the entries which are important for our calculations.

What is now different are the pseudovacua. Indeed, we want to show that in
the coordinate representation the pseudovacua are given by the following element of ${\cal T}({\cal H})$:
\begin{equation}
{\bf \Omega} (x_1, \ldots , x_N)=\prod _{i=1}^{N/4}  \bar f(x_{4i-1}-x_{4i})
f(x_{4i-3}-x_{4i-2}) \,  \, \delta \left (\sum _{i=1}^{N/4}
(x_{4i-3}-x_{4i-1})\right ) \, ,
\label {Pse}
\end{equation}
where the function $f(x)$ is a solution of (\ref{thepro}) and $\bar f(x)$ a
solution of (\ref{rthepro}).

Let us prove this statement in some steps. These annihilation properties,
derived from (\ref{coorep3}, \ref{thepro}, \ref{rcoorep3}, \ref{rthepro}),
are crucial: \begin{equation} [\bar F_{21}^{(4i,4i-1)}(\lambda ){\bf \Omega}
](x_1, \ldots , x_N)=0 \quad , \quad  [F_{21}^{(4i-2,4i-3)}(\lambda ){\bf
\Omega}](x_1, \ldots , x_N)=0 \, .  \label{Offdia}
\end{equation}
Then, we consider the expressions of ${\bf A(\la;\mu)}$, ${\bf D(\la;\mu)}$,
${\bf C(\la;\mu)}$ in terms of the entries of the fused Lax operators. For
instance, if $N=8$, we have (for conciseness we omit that $F$'s depend on the
combination $\mu ^{\frac {1}{2}}\lambda $, instead $\bar F$'s on the
combination $\mu ^{\frac {1}{2}}/\lambda $):
\begin{eqnarray}
&& {\bf A}(\lambda;\mu)=\left [ \bar F_{11}^{(8,7)}F_{11}^{(6,5)}+\bar
F_{12}^{(8,7)}F_{21}^{(6,5)}\right ] \left [ \bar
F_{11}^{(4,3)}F_{11}^{(2,1)}+ \bar F_{12}^{(4,3)}F_{21}^{(2,1)}\right ] +
\nonumber \\
&& + \left [ \bar F_{11}^{(8,7)}F_{12}^{(6,5)}+\bar
F_{12}^{(8,7)}F_{22}^{(6,5)}\right ]\left [ \bar
F_{21}^{(4,3)}F_{11}^{(2,1)}+\bar F_{22}^{(4,3)}F_{21}^{(2,1)}\right ] \, ,
\nonumber \\
&& {\bf D}(\lambda;\mu)=\left [ \bar
F_{21}^{(8,7)}F_{11}^{(6,5)}+\bar F_{22}^{(8,7)}F_{21}^{(6,5)}\right ]\left
[\bar F_{11}^{(4,3)} F_{12}^{(2,1)}+\bar F_{12}^{(4,3)}F_{22}^{(2,1)}\right ]
+\nonumber \\ && + \left [ \bar F_{21}^{(8,7)}F_{12}^{(6,5)}+\bar
F_{22}^{(8,7)}F_{22}^{(6,5)}\right ]\left [ \bar
F_{21}^{(4,3)}F_{12}^{(2,1)}+\bar F_{22}^{(4,3)}F_{22}^{(2,1)}\right ] \, ,
\label{N8} \\
&& {\bf C}(\lambda;\mu)=\left [ \bar
F_{21}^{(8,7)}F_{11}^{(6,5)}+\bar F_{22}^{(8,7)}F_{21}^{(6,5)}\right ]\left [
\bar F_{11}^{(4,3)}F_{11}^{(2,1)}+\bar F_{12}^{(4,3)}F_{21}^{(2,1)}\right ]
+\nonumber \\ && + \left [ \bar F_{21}^{(8,7)}F_{12}^{(6,5)}+\bar
F_{22}^{(8,7)}F_{22}^{(6,5)}\right ]\left [ \bar
F_{21}^{(4,3)}F_{11}^{(2,1)}+\bar F_{22}^{(4,3)}F_{21}^{(2,1)}\right ] \, .
\nonumber
\end{eqnarray}
Hence, it is crucial that $F_{ij}^{(k,k-1)}(\la)$ and $\bar
F_{i^{\prime}j^{\prime}}^{(k^{\prime},k^{\prime}-1)}(\la^{\prime})$
($k,\,k^\prime\in 2{\Bbb N}$; $\,2\leq k,\, k^\prime \leq
N$) -- not necessarily in a representation -- satisfy, in addition to the
previous ones (\ref{brafus1}-\ref{brafus4}) and
(\ref{rbrafus1}-\ref{rbrafus4}), mixed exchange relations, following directly
from the $W$'s exchange algebra:

\begin{itemize}

\item exchange $(21)$-$(11)$
\ba
F_{21}^{(4i+2,4i+1)}(\la) \bar
F_{11}^{(4i,4i-1)}(\la^{\prime})&=&q^{\frac{1}{2}}  \bar F_{11}^{(4i,4i-1)}(\la^{\prime})
F_{21}^{(4i+2,4i+1)}(\la) \, , \nn \\
\bar F_{21}^{(4i,4i-1)}(\la^{\prime})
F_{11}^{(4i-2,4i-3)}(\la)&=&q^{-\frac{1}{2}}  F_{11}^{(4i-2,4i-3)}(\la) \bar
F_{21}^{(4i,4i-1)}(\la^{\prime}) \, , \nn \\
\bar F_{21}^{(N,N-1)}(\la^{\prime}) F_{11}^{(2,1)}(\la)&=&
q^{\frac{1}{2}} F_{11}^{(2,1)}(\la) \bar F_{21}^{(N,N-1)}(\la^{\prime}) \, ;
\label{Brafus1}
\ea

\item exchange $(21)$-$(12)$
\ba
F_{21}^{(4i+2,4i+1)}(\la) \bar F_{12}^{(4i,4i-1)}(\la^{\prime})&=&q^{\frac {1}{2}}
 \bar F_{12}^{(4i,4i-1)}(\la^{\prime}) F_{21}^{(4i+2,4i+1)}(\la) \, , \nn \\
\bar
F_{21}^{(4i,4i-1)}(\la^{\prime})F_{12}^{(4i-2,4i-3)}(\la)&=&q^{-\frac{1}{2}}
F_{12}^{(4i-2,4i-3)}(\la) \bar F_{21}^{(4i,4i-1)}(\la^{\prime}) \, , \nn \\
\bar F_{21}^{(N,N-1)}(\la^{\prime}) F_{12}^{(2,1)}(\la)&=&q^{-\frac{1}{2}}
F_{12}^{(2,1)}(\la) \bar F_{21}^{(N,N-1)} (\la^{\prime})\, ;
\label{Brafus2}
\ea

\item exchange $(21)$-$(22)$
\ba
F_{21}^{(4i+2,4i+1)}(\la) \bar F_{22}^{(4i,4i-1)}(\la^{\prime})&=&q^{-\frac{1}{2}}
\bar F_{22}^{(4i,4i-1)}(\la^{\prime}) F_{21}^{(4i+2,4i+1)}(\la) \, , \nn \\
\bar F_{21}^{(4i,4i-1)}(\la^{\prime})
F_{22}^{(4i-2,4i-3)}(\la)&=&q^{\frac{1}{2}}  F_{22}^{(4i-2,4i-3)}(\la) \bar
F_{21}^{(4i,4i-1)}(\la^{\prime}) \, , \nn \\
\bar F_{21}^{(N,N-1)}(\la^{\prime}) F_{22}^{(2,1)}(\la)&=&q^{-\frac{1}{2}}
F_{22}^{(2,1)}(\la) \bar F_{21}^{(N,N-1)}(\la^{\prime}) \, ;
\label{Brafus3}
\ea

\item commutation if $2<|k-k^\prime|<N-2$
\be
[\bar F_{ij}^{(k,k-1)}(\la) \, , \, F_{i^\prime j^\prime}^{(k^\prime ,
k^\prime -1)} (\la ^\prime)]=0  \, .
\label{Brafus4}
\ee

\end{itemize}

Indeed, after iterated use of the exchange properties
(\ref{Brafus1}-\ref{Brafus4}), we can accumulate all the factors $\bar
F_{21}^{(4i,4i-1)}, F_{21}^{(4i-2,4i-3)}$ to the right of the addenda in
expressions of ${\bf A}$, ${\bf D}$, ${\bf C}$. From the form
of these (see, for example, formul\ae \, (\ref {N8}) in the case $N=8$) and
from annihilation properties (\ref{Offdia}) it then follows:
\begin{eqnarray}
{\bf A}(\lambda ; \mu)\, {\bf
\Omega} &=& \prod _{i=1}^{\stackrel {N/4}{\leftarrow}} \bar F_{11}^{(4i,4i-1)}
\left ( \frac {\mu ^{1/2}}{\lambda }\right ) F_{11}^{(4i-2,4i-3)}  (\mu
^{1/2}\lambda )\,  {\bf \Omega}  \, , \nn \\
{\bf D}(\lambda ;\mu)\, {\bf \Omega} &=&  \prod _{i=1}^{\stackrel
{N/4}{\leftarrow}} \bar F_{22}^{(4i,4i-1)} \left ( \frac {\mu ^{1/2}}{\lambda
}\right ) F_{22}^{(4i-2,4i-3)} (\mu ^{1/2}\lambda )\,  {\bf \Omega} \, , \nn \\
\nn \\
{\bf C}(\lambda ;\mu )\, {\bf \Omega} &=& 0 \, .
\label{COM}
\end{eqnarray}
We have already proved part of the statement in (\ref{COM}) and we complete
through finding the eigenvalues of ${\bf A}$ and ${\bf D}$ over ${\bf
\Omega}$ in the following theorem and corollary.

\begin{theorem}
\label {nonconf1}
The action of the ordered products of the operators ${\bf F}_{11}^i$ and ${\bf
F}_{22}^i$ -- defined by
\ba
{\bf F}_{11}^i(\lambda;\mu) &\equiv& \bar F_{11}^{(4i,4i-1)}
\left ( \frac {\mu ^{1/2}}{\lambda }\right ) F_{11}^{(4i-2,4i-3)}(\mu^{1/2}
\lambda )\, , \\
\label{defF11}
{\bf F}_{22}^i(\lambda;\mu) &\equiv& \bar F_{22}^{(4i,4i-1)} \left(\frac{\mu^{1/2}}
{\lambda }\right ) F_{22}^{(4i-2,4i-3)} (\mu^{1/2}\lambda )\, ,
\label{defF22}
\ea
-- on the states (\ref{Pse}) is the following ($1\leq i \leq N/4$):
\begin{eqnarray}
&&\Bigl
[\prod _{j=1}^{\stackrel {i}{\leftarrow}} {\bf F}_{11}^j(\lambda; \mu){\bf
\Omega}\Bigr ](x_1,\ldots ,x_N)=\nonumber \\ && q^{-\frac {1}{2}}
e^{-i(x_{4i+1}+2{\sum \limits _{j=1}^{2i-1}}\, (-)^jx_{2j+1}+x_1)}(1-\Delta ^2
\mu \lambda ^2 q^{-1})^i\left (1-\Delta ^2 \frac {\mu}{\lambda ^2} q\right)^i
{\bf \Omega} (x_1,\ldots ,x_N) \, , \nonumber \\ &&\Bigl [\prod
_{j=1}^{\stackrel {i}{\leftarrow}} {\bf F}_{22}^j(\lambda; \mu){\bf \Omega}\Bigr
](x_1,\ldots ,x_N)=\nonumber \\ && q^{-\frac {1}{2}} e^{i(x_{4i+1}+2\sum
\limits _{j=1}^{2i-1}\, (-)^jx_{2j+1}+x_1)}(1-\Delta ^2 \mu \lambda ^2
q)^i\left (1-\Delta ^2 \frac {\mu }{\lambda ^2} q^{-1}\right)^i {\bf \Omega}
(x_1,\ldots ,x_N)\, . \nonumber  \end{eqnarray} \end{theorem} {\bf Proof}: We
show by induction the first formula. For $i=1$ we have, using formul\ae \,
(\ref {coorep1}, \ref {rcoorep1}):  \begin{eqnarray} && [{\bf
F}_{11}^1(\lambda;\mu){\bf \Omega}](x_1,\ldots ,x_N)=e^{-i\left(x_{5}-2x_{3}+x_{1}
-\frac {\pi \beta ^2}{2}\right )}{\bf \Omega} (x_{1}^+, x_{2}^+, x_{3}^+,
x_{4}^+,x_5,\ldots , x_N) + \nonumber \\ && + \Delta ^2 \mu \lambda ^2 q^{-1}
e^{-i\left(x_{5}-2x_{3}+2x_{2}-x_{1} -\frac {\pi \beta ^2}{2}\right )}{\bf
\Omega} (x_{1}^+, x_{2}^-, x_{3}^+, x_{4}^+,x_5,\ldots , x_N) + \nonumber \\
&&+ \Delta ^2 \frac {\mu }{\lambda ^2 }q e^{-i\left(x_{5}-2x_{4}+x_{1} -\frac
{\pi \beta ^2}{2}\right )}{\bf \Omega} (x_{1}^+, x_{2}^+, x_{3}^+,
x_{4}^-,x_5,\ldots , x_N) + \nonumber \\ && + \Delta ^4 \mu ^2
e^{-i\left(x_{5}-2x_{4}+2x_{2}-x_{1} -\frac {\pi \beta ^2}{2}\right )}{\bf
\Omega} (x_{1}^+, x_{2}^-, x_{3}^+, x_{4}^-,x_5,\ldots , x_N) \, .
\label{interm} \end{eqnarray} Now we remark that ${\bf \Omega}
(x_{1}^+,x_{2}^+, x_{3}^+, x_{4}^+,x_5,\ldots , x_N)= {\bf \Omega} (x_1,
\ldots , x_N)$ and that the use of the shift  properties (\ref{thepro},
\ref{rthepro}) for the functions $f$, $\bar f$ contained  in (\ref {Pse})
gives
\begin{eqnarray}
&& {\bf \Omega}(x_{1}^+,x_{2}^-, x_{3}^+, x_{4}^+,x_5,\ldots , x_N)=-e^{-2i(x_1-x_2)}
{\bf \Omega} (x_{1}, \ldots , x_N) \, , \nonumber \\
&& {\bf \Omega} (x_{1}^+,x_{2}^+, x_{3}^+, x_{4}^-,x_5,\ldots , x_N)=-e^{2i(x_3-x_4)}
{\bf \Omega} (x_{1},\ldots , x_N) \, , \nonumber \\
&& {\bf \Omega} (x_{1}^+,x_{2}^-, x_{3}^+, x_{4}^-,x_5,\ldots , x_N)=e^{-2i(x_1-x_2)}
e^{2i(x_3-x_4)}{\bf \Omega} (x_{1}, \ldots , x_N) \,
, \nonumber
\end{eqnarray}
because the shifts in the variables $x_{1}, ..., x_{4}$ do not affect
the delta function contained in (\ref {Pse}). Therefore, all the terms in
(\ref {interm}) are proportional and the final result is:
\begin{eqnarray}
&[{\bf F}_{11}^1(\lambda;\mu){\bf \Omega} ](x_1,\ldots ,x_N)=q^{-\frac {1}{2}} e^{-i(x_{5}-2x_{3}+x_1)}\left (1-\frac {\Delta ^2\mu \lambda ^2}{q}\right)\left (1-\frac {\Delta ^2 \mu q}{\lambda ^2} \right) {\bf \Omega} (x_1,\ldots ,x_N) , \nonumber &
\end{eqnarray}
which is the first formula of Theorem \ref {nonconf1} for $i=1$.

For $2\leq i\leq N/4$ we have from (\ref {coorep1}, \ref {rcoorep1}):
\begin{eqnarray}
&&\Bigl [\prod _{j=1}^{\stackrel {i}{\leftarrow}} {\bf F}_{11}^j(\lambda;\mu){\bf \Omega}\Bigr ](x_1,\ldots ,x_N)= q^{-\frac {1}{2}} e^{-ix_{4i+1}}
\Bigl \{ e^{i(2x_{4i-1}-x_{4i-3} )} \cdot
 \nonumber \\
&& \cdot  \Bigl [\prod _{j=1}^{\stackrel {i-1}{\leftarrow}} {\bf F}_{11}^j(\lambda;\mu){\bf \Omega}\Bigr ](x_1,\ldots ,x_{4i-4},  x_{4i-3}^+,x_{4i-2}^+,x_{4i-1}^+,x_{4i}^+,x_{4i+1}, \ldots , x_N)+
\nonumber \\
&&+ \frac {\Delta ^2 \mu \lambda ^2}{q} e^{i(2x_{4i-1}-2x_{4i-2}+x_{4i-3})} \cdot \nonumber \\
&& \cdot
\Bigl [\prod _{j=1}^{\stackrel {i-1}{\leftarrow}} {\bf F}_{11}^j(\lambda;\mu){\bf \Omega}\Bigr ](x_1,\ldots ,x_{4i-4},  x_{4i-3}^+,x_{4i-2}^-,x_{4i-1}^+,x_{4i}^+,x_{4i+1}, \ldots , x_N)+
\frac {\Delta ^2 \mu q }{\lambda ^2} \cdot \nonumber \\
&& \cdot
 e^{i(2x_{4i}-x_{4i-3})}\Bigl [\prod _{j=1}^{\stackrel {i-1}{\leftarrow}} {\bf F}_{11}^j(\lambda;\mu){\bf \Omega}\Bigr ](x_1,\ldots ,x_{4i-4},  x_{4i-3}^+,x_{4i-2}^+,x_{4i-1}^+,x_{4i}^-,x_{4i+1}, \ldots , x_N)+ \nonumber \\
&& + \Delta ^4 \mu ^2  e^{i(2x_{4i}-2x_{4i-2}+x_{4i-3})} \cdot \nonumber \\
&&\cdot
\Bigl [\prod _{j=1}^{\stackrel {i-1}{\leftarrow}} {\bf F}_{11}^j(\lambda;\mu){\bf \Omega}\Bigr ](x_1,\ldots ,x_{4i-4},  x_{4i-3}^+,x_{4i-2}^-,x_{4i-1}^+,x_{4i}^-,x_{4i+1},\ldots , x_N) \Bigr \} \, .
\nonumber
\end{eqnarray}

Using the inductive hypothesis, we get:
\begin{eqnarray}
&& \Bigl [\prod _{j=1}^{\stackrel {i}{\leftarrow}} {\bf F}_{11}^j(\lambda;\mu){\bf \Omega}\Bigr ](x_1,\ldots ,x_N)= \nonumber \\
&& =q^{-1} \Bigl \{ e^{-i(x_{4i+1}-2x_{4i-1}+x_{4i-3} )}
e^{-i(x^+_{4i-3}+2\sum\limits _{j=1}^{2i-3}(-)^jx_{2j+1}+x_1)}
 (1-\Delta ^2 \mu \lambda ^2 q^{-1})^{i-1} \cdot
\nonumber \\
&&\cdot\left (1-\Delta ^2 \frac {\mu}{\lambda ^2} q\right)^{i-1} {\bf \Omega}(x_1,\ldots ,x_{4i-4},  x_{4i-3}^+,x_{4i-2}^+,x_{4i-1}^+,x_{4i}^+,x_{4i+1}, \ldots , x_N)+ \nonumber \\
&&+ \frac {\Delta ^2 \mu \lambda ^2}{q}  e^{-i(x_{4i+1}-2x_{4i-1}+2x_{4i-2}-x_{4i-3})}
e^{-i(x^+_{4i-3}+2\sum \limits_{j=1}^{2i-3}(-)^jx_{2j+1}+x_1)}
 (1-\Delta ^2 \mu \lambda ^2 q^{-1})^{i-1} \cdot
\nonumber \\
&&\cdot\left (1-\Delta ^2 \frac {\mu}{\lambda ^2} q\right)^{i-1} {\bf \Omega}(x_1,\ldots ,x_{4i-4},x_{4i-3}^+,x_{4i-2}^-,x_{4i-1}^+,x_{4i}^+, x_{4i+1},\ldots , x_N)+ \nonumber \\
&& + \frac {\Delta ^2 \mu q}{\lambda ^2}
 e^{-i(x_{4i+1}-2x_{4i}+x_{4i-3})}e^{-i(x^+_{4i-3}+2\sum \limits _{j=1}^{2i-3}(-)^jx_{2j+1}+x_1)}
 (1-\Delta ^2 \mu \lambda ^2 q^{-1})^{i-1} \cdot
\nonumber \\
&&\cdot \left (1-\Delta ^2 \frac {\mu}{\lambda ^2} q\right)^{i-1} {\bf \Omega} (x_1,\ldots, x_{4i-4}, x_{4i-3}^+,  ,x_{4i-2}^+,x_{4i-1}^+,x_{4i}^-,x_{4i+1}, \ldots , x_N)+ \nonumber \\
&& + \Delta ^4 \mu ^2  e^{-i(x_{4i+1}-2x_{4i}+2x_{4i-2}-x_{4i-3})}
e^{-i(x^+_{4i-3}+2\sum\limits _{j=1}^{2i-3}(-)^jx_{2j+1}+x_1)}
(1-\Delta ^2 \mu \lambda ^2 q^{-1})^{i-1}  \cdot \nonumber \\
&& \cdot  \left (1-\Delta ^2 \frac {\mu}{\lambda ^2} q\right)^{i-1} {\bf \Omega} (x_1,\ldots,x_{4i-4}, x_{4i-3}^+,x_{4i-2}^-,x_{4i-1}^+,x_{4i}^-, x_{4i+1}, \ldots , x_N) \Bigr \} \, . \label{Interm}
\end{eqnarray}
As in the case $i=1$ we have:
\begin{equation}
{\bf \Omega} (x_1,\ldots ,x_{4i-4},  x_{4i-3}^+,x_{4i-2}^+,x_{4i-1}^+,x_{4i}^+,x_{4i+1}, \ldots , x_N)={\bf \Omega} (x_1, \ldots , x_N) \, , \nonumber
\end{equation}
and the use of property (\ref{thepro}, \ref{rthepro}) for the functions $f$, $\bar f$ contained
in (\ref {Pse}) gives:
\begin{eqnarray}
&& {\bf \Omega}(x_1, .. ,x_{4i-4},x_{4i-3}^+,x_{4i-2}^-,x_{4i-1}^+,x_{4i}^+,x_{4i+1}, .. , x_N)=-e^{-2i(x_{4i-3}-x_{4i-2})} {\bf \Omega}(x_1, .. ,x_N) \, , \nonumber \\
&& {\bf \Omega}(x_1, .. ,x_{4i-4},x_{4i-3}^+,x_{4i-2}^+,x_{4i-1}^+,x_{4i}^-, x_{4i+1}, .. , x_N)=-e^{2i(x_{4i-1}-x_{4i})} {\bf \Omega}(x_1, .. ,x_N)
\, ,  \nonumber \\
&& {\bf \Omega}(x_1, .. ,x_{4i-4},x_{4i-3}^+,x_{4i-2}^-,x_{4i-1}^+,x_{4i}^-, x_{4i+1}, .. , x_N)= \nonumber \\
&& \hspace {7cm}= e^{-2i(x_{4i-3}-x_{4i-2})}e^{2i(x_{4i-1}-x_{4i})} {\bf \Omega} (x_1, .. ,x_N)
\, , \nonumber
\end{eqnarray}
because the shifts in the variables $x_{4i-3}, ..., x_{4i}$ do not affect
the delta function contained in (\ref {Pse}). Hence, all the terms in (\ref
{Interm}) are proportional to ${\bf \Omega}$ and the final result is:
\begin{eqnarray} &&\Bigl [\prod _{j=1}^{\stackrel {i}{\leftarrow}} {\bf
F}_{11}^j(\lambda;\mu){\bf \Omega}\Bigr ](x_1,\ldots ,x_N)=\nonumber \\ && =
q^{-\frac {1}{2}} e^{-i(x_{4i+1}+2\sum\limits
_{j=1}^{2i-1}(-)^jx_{2j+1}+x_1)}(1-\Delta ^2 \mu \lambda ^2 q^{-1})^i\left
(1-\Delta ^2 \frac {\mu}{\lambda ^2} q\right)^i {\bf \Omega} (x_1,\ldots ,x_N)
\, , \nonumber \\ \end{eqnarray} which is the first formula of Theorem \ref
{nonconf1}.

The proof for ${\bf F}_{22}$ elements follows the same lines and we do not
write it.
\finproof

\medskip

\begin{corollary}
\label {cornonconf}
The states (\ref {Pse}) are eigenvectors of the elements ${\bf A}(\lambda;\mu)$
and ${\bf D}(\lambda;\mu)$ of the monodromy matrix (\ref {MON}). The
corresponding common eigenvalues are given by the following formul\ae:
\begin{eqnarray} {\bf A}(\lambda;\mu){\bf \Omega} &=& q^{-\frac {1}{2}}
(1-\Delta ^2 \mu \lambda ^2 q^{-1})^{N/4}\left (1-\Delta ^2 \frac
{\mu}{\lambda ^2} q\right)^{N/4} {\bf \Omega} \, , \label{AOM} \\
{\bf D}(\lambda;\mu){\bf \Omega} &=& q^{-\frac {1}{2}}
(1-\Delta ^2 \mu \lambda ^2 q)^{N/4}\left (1-\Delta ^2 \frac {\mu}{\lambda ^2}
q^{-1}\right)^{N/4} {\bf \Omega} \, .
\label{DOM}
\end{eqnarray}
\end{corollary}
{\bf Proof}:
We apply Theorem \ref {nonconf1} for $i=N/4$ and remember that the variable
$x_{N+1}$, appearing in formul\ae \, of Theorem  \ref {nonconf1} for $i=N/4$,
must be read as $x_1$. Hence, we have that the exponents in the second factors
in the right hand sides of formul\ae \, of Theorem  \ref {nonconf1} are
proportional to $\sum\limits _{i=1}^{N/4} (x_{4i-3}-x_{4i-1})$: therefore they
can be put equal to zero because of the delta function $\delta \left
(\sum\limits _{i=1}^{N/4} (x_{4i-3}-x_{4i-1})\right )$ in the definition (\ref
{Pse}) of ${\bf \Omega}$. In such a way we obtain formul\ae \, (\ref {AOM},
\ref {DOM}). \finproof

\medskip

Eventually, formul\ae \, (\ref{COM}, \ref {AOM}, \ref {DOM}) show that (\ref{Pse})
are pseudovacuum states for the monodromy matrix (\ref{MON}) with the same
${\bf A}(\lambda;\mu)$ and ${\bf D}(\lambda;\mu)$ eigenvalues, respectively,
for any $f(x)$ solution of (\ref{thepro}) and any $\bar f(x)$ solution
(\ref{rthepro}). Nevertheless, {\it a fortiori} the pseudovacua are
non-ultralocal in this off-critical case.

In order to write down the BE's, we remark that from (\ref{pmon2})
it follows that monodromy matrix (\ref {MON}) satisfy the exchange relations
(\ref{rll}) in which $Z^{-1}_{ab}$ is replaced by $Z_{ab}$. Hence, the
braided exchange rules between ${\bf B}(\la ^\prime;\mu)$ and ${\bf
A}(\la;\mu)$, ${\bf D}(\la;\mu)$ are respectively (we suppress the dependence
on $\mu$ for reasons of conciseness):
\begin{eqnarray}
{\bf A}(\lambda){\bf B}(\lambda
^\prime)&=&\frac {q}{a\left(\frac   {\lambda ^\prime}{\lambda}\right)}{\bf
B}(\lambda ^\prime){\bf A}(\lambda) - {q}\frac {b\left(\frac   {\lambda
^\prime}{\lambda}\right)}{a\left(\frac   {\lambda
^\prime}{\lambda}\right)}{\bf B}(\lambda){\bf A}(\lambda ^\prime) \, , \label
{Exch1} \\ {\bf D}(\lambda){\bf B}(\lambda ^\prime)&=&\frac
{q^{-1}}{a\left(\frac   {\lambda}{\lambda ^\prime}\right)}{\bf B}(\lambda
^\prime){\bf D}(\lambda) - {q^{-1}}\frac {b\left(\frac {\lambda}{\lambda
^\prime}\right)}{a\left(\frac   {\lambda}{\lambda ^\prime}\right)}{\bf
B}(\lambda){\bf D}(\lambda ^\prime) \, .
\label{Exch2}
\end{eqnarray}
As in the previous Section, the states
\begin{equation}
{\bf \Psi} (\lambda _1, \ldots, \lambda _l)=\prod _{r=1}^l{\bf B}(\lambda_r){\bf \Omega} \,
\label{Psi}
\end{equation}
are eigenstates of the transfer matrix ${\bf T}(\lambda)={\bf
A}(\lambda)+{\bf D}(\lambda)$ (Bethe states) only if the set of complex numbers
$\{ \lambda _1, \ldots , \lambda _l \} $ (Bethe roots) satisfy the following
Bethe Equations (BE's):
\begin{equation} q^{2l}\prod _{{\stackrel{r=1}{r\not= s}}}^l \frac
{q\lambda _r^2-q^{-1}\lambda   _s^2}{q^{-1}\lambda _r^2-q\lambda _s^2}=\left [
\frac {(1- \Delta ^2 \mu \lambda _s ^2  q)\left(1-\Delta ^2\frac
  {\mu}{\lambda _s ^2}q^{-1}\right)}
{(1-\Delta ^2 \mu \lambda _s ^2 q^{-1})\left(1-\Delta ^2\frac
  {\mu}{\lambda _s^2}q\right)}\right]^{N/4} \, .
\label{Bethe2}
\end{equation}
It is useful to rewrite (\ref {Bethe2}) in a trigonometric form. Let us
define the new variables $\Theta$, $\alpha$ and $\alpha_r$:
\begin{equation}
\Delta ^2 \mu \equiv e^{-2\Theta} \quad , \quad \lambda \equiv e^{\alpha}
\quad , \quad \lambda_r \equiv e^{\alpha_r} \quad .
\label{trigon}
\end{equation}
In terms of these variables the BE's (\ref{Bethe2}) are
($q=e^{-i\pi \beta ^2}$):
\begin{equation}
e^{-2i\pi \beta ^2 l}\prod _{{\stackrel{r=1}{r\not= s}}}^l \frac {\sinh (\alpha _s -\alpha _r +i\pi \beta ^2)}{\sinh (\alpha _s -\alpha _r -i\pi \beta ^2)}=\left [\frac
{\sinh \left (\alpha _s +\Theta -\frac {i\pi \beta ^2}{2}\right)\sinh \left(\alpha _s -\Theta -\frac {i\pi\beta ^2}{2}\right)}
{\sinh \left (\alpha _s +\Theta +\frac {i\pi \beta ^2}{2}\right)\sinh
\left (\alpha _s -\Theta +\frac {i\pi \beta ^2}{2}\right)}
\right]^{N/4} \, .
\label{Bethe3}
\end{equation}
Finally, from equations (\ref {Exch1},\ref{Exch2}) and from (\ref{AOM}) and
(\ref{DOM}) it  follows that the eigenvalues ${\bf \Lambda}(\lambda ,
\{\lambda _r\})$ of the transfer matrix ${\bf T}(\lambda)$ on the Bethe states
(\ref{Psi},\ref{Bethe2}) are:
\begin{eqnarray}
{\bf \Lambda}(\lambda , \{ \lambda _r\})&=&
q^{l}\prod _{r=1}^l
\frac {q^{-1}\lambda
  ^2-q\lambda _r^2}{\lambda ^2-\lambda
  _r^2}q^{-\frac {1}{2}}[(1-\Delta ^2\mu \lambda ^2q^{-1})(1-\Delta ^2\mu \lambda ^{-2}q)]^{N/4}+\nonumber \\
 &+&q^{-l}\prod _{r=1}^l
\frac {q\lambda ^2-q^{-1}\lambda _r^2}{\lambda ^2-\lambda
  _r^2}q^{-\frac {1}{2}}[(1-\Delta ^2\mu \lambda ^2q)(1-\Delta ^2\mu \lambda ^{-2}q^{-1})]^{N/4} \, . \label {RT}
\end{eqnarray}
In addition, it is useful to write also the eigenvalues of the transfer matrix
in a trigonometric form. After inserting (\ref{trigon}) in (\ref{RT}), we
obtain:
\begin{eqnarray}
&e^{-\frac {i\pi \beta ^2}{2}+\frac {\Theta N}{2}}
{\bf \Lambda}(\alpha , \{ \alpha _r \})=& \label {RTtrig} \\
&=e^{-i\pi \beta^2 l} \prod\limits _{r=1}^l \frac {\sinh (\alpha -\alpha _r+i\pi \beta ^2)}{\sinh (\alpha -\alpha _r) } \left [ \sinh \left ( \Theta -\alpha -\frac {i\pi \beta ^2}{2} \right )
\sinh \left (\Theta +\alpha +\frac {i\pi \beta ^2}{2} \right ) \right ]^{N/4}
+&\nonumber \\
&+ e^{i\pi \beta ^2 l} \prod\limits _{r=1}^l \frac {\sinh (\alpha -\alpha _r-i\pi \beta ^2)}{\sinh (\alpha -\alpha _r) } \left [ \sinh \left ( \Theta -\alpha +\frac {i\pi \beta ^2}{2} \right )
\sinh \left (\Theta +\alpha - \frac {i\pi \beta ^2}{2} \right ) \right ]^{N/4}
.& \nonumber
\end{eqnarray}

\medskip

Now, for completeness, we illustrate the main results regarding the other
choice of off-critical monodromy matrix (\ref {Leftmon}). The calculations for
obtaining
\begin{itemize}
\item the pseudovacua,
\item the Bethe states and the the Bethe Equations,
\item the eigenvalues of the transfer matrix
\end{itemize}
have been carried out in a way parallel to that performed in case
(\ref{Rightmon}). In what follows, we summarize only the final results.

The pseudovacua in the coordinate representation are given by
\begin{equation}
{\bf \Omega}^\prime (x_1, \ldots , x_N)=\prod _{i=1}^{N/4} f(x_{4i-1}-x_{4i})
\bar f(x_{4i-3}-x_{4i-2})\,  \, \delta \left (\sum _{i=1}^{N/4}
(x_{4i-3}-x_{4i-1})\right ) \, .
\nonumber
\end{equation}

The Bethe states are:
\be
{\bf \Psi}^{\prime}(\lambda _1, \ldots, \lambda _l)=\prod _{r=1}^l{\bf
B^{\prime}}(\lambda_r){\bf \Omega}^{\prime} \, ,
\label{Psi'}
\ee
in addition to the BE's
\begin{equation}
q^{-2l}\prod _{\stackrel {r=1}{r\not=s}}^l \frac {q\lambda _r^2-q^{-1}\lambda
  _s^2}{q^{-1}\lambda _r^2-q\lambda _s^2}=\left [
\frac {(1- \Delta ^2 \mu \lambda _s ^2  q)\left(1-\Delta ^2\frac
  {\mu}{\lambda _s ^2}q^{-1}\right)}
{(1-\Delta ^2 \mu \lambda _s ^2 q^{-1})\left(1-\Delta ^2\frac
  {\mu}{\lambda _s^2}q\right)}\right]^{N/4}
\, , \label{Bethe4}
\end{equation}
or in trigonometric form
\begin{equation}
e^{2i\pi \beta ^2 l}\prod _{{\stackrel{r=1}{r\not= s}}}^l \frac {\sinh (\alpha _s -\alpha _r +i\pi \beta ^2)}{\sinh (\alpha _s -\alpha _r -i\pi \beta ^2)}=\left [\frac
{\sinh \left (\alpha _s +\Theta -\frac {i\pi \beta ^2}{2}\right)\sinh \left(\alpha _s -\Theta -\frac {i\pi\beta ^2}{2}\right)}
{\sinh \left (\alpha _s +\Theta +\frac {i\pi \beta ^2}{2}\right)\sinh
\left (\alpha _s -\Theta +\frac {i\pi \beta ^2}{2}\right)}
\right]^{N/4} \, .
\label{Bethe5}
\end{equation}

The eigenvalues of the transfer matrix are:
\begin{eqnarray}
{\bf \Lambda}^\prime (\lambda , \{ \lambda _r\})&=&
q^{-l}\prod _{r=1}^l
\frac {q^{-1}\lambda^2-q\lambda_r^2}{\lambda ^2-\lambda_r^2}
q^{\frac {1}{2}}[(1-\Delta ^2\mu \lambda ^2q^{-1})(1-\Delta ^2\mu \lambda
^{-2}q)]^{N/4}+\nonumber \\  &+&q^{l}\prod _{r=1}^l
\frac {q\lambda ^2-q^{-1}\lambda _r^2}{\lambda ^2-\lambda_r^2}q^{\frac{1}{2}}
[(1-\Delta ^2\mu \lambda ^2q)(1-\Delta ^2\mu \lambda ^{-2}q^{-1})]^{N/4} \, ,
\label {LT}
\end{eqnarray}
or in trigonometric form
\begin{eqnarray}
&e^{\frac {i\pi \beta ^2}{2}+\frac {\Theta N}{2}}{\bf \Lambda}^\prime (\alpha , \{ \alpha _r \})=& \label {LTtrig} \\
&= e^{i\pi \beta^2 l} \prod\limits _{r=1}^l \frac {\sinh (\alpha -\alpha _r+i\pi \beta ^2)}{\sinh (\alpha -\alpha _r) } \left [ \sinh \left (\Theta -\alpha -\frac {i\pi \beta ^2}{2} \right )
\sinh \left (\Theta +\alpha +\frac {i\pi \beta ^2}{2} \right ) \right ]^{N/4}
 +&\nonumber \\
&+ e^{-i\pi \beta ^2 l} \prod\limits _{r=1}^l \frac {\sinh (\alpha -\alpha _r-i\pi \beta ^2)}{\sinh (\alpha -\alpha _r) } \left [ \sinh \left (\Theta -\alpha +\frac {i\pi \beta ^2}{2} \right )
\sinh \left (\Theta +\alpha - \frac {i\pi \beta ^2}{2} \right ) \right ]^{N/4}.& \nonumber
\end{eqnarray}

In this section we have calculated the eigenvalues of the two lattice transfer
matrices associated to the monodromy matrices (\ref{Rightmon}) and
(\ref{Leftmon}). We will show in next Section that these eigenvalues in the
{\it limit} $\mu \rightarrow 0$ reduce to the conformal right and left
ones respectively. Consequently, this reinforce our idea that the monodromy matrices
(\ref{Rightmon}) and (\ref{Leftmon}) will describe, after (cylinder) continuum
limit, {\it a sort of perturbation} from CFT. We will come back rigorously on
the nature of these theories in a future publication \cite{35}.

\section{Conformal limits of the off-critical transfer matrix eigenvalues.}
\setcounter{equation}{0}
In this section we want to show that, after suitable rescaling of the spectral
parameter and of the Bethe roots, in the limit $\mu \rightarrow 0$ the
eigenvalues of the off-critical transfer matrices (\ref{RT}) and (\ref{LT}) are
proportional respectively to the eigenvalues of the right and left conformal
transfer matrices (\ref{rT}) and (\ref{T}).

\medskip

Indeed, let us consider the eigenvalue (\ref{RT}) and let us calculate the
limit:
\begin{eqnarray}
\lim _{\mu \rightarrow 0}
{\bf \Lambda}(\lambda \mu ^{1/2}, \{ \lambda _r \mu ^{1/2} \} )&=&
q^{l}\prod _{r=1}^l
\frac {q^{-1}\lambda
  ^2-q\lambda _r^2}{\lambda ^2-\lambda
  _r^2}q^{-\frac {1}{2}}(1-\Delta ^2 \lambda ^{-2}q)^{N/4}+\nonumber \\
 &+& q^{-l}\prod _{r=1}^l
\frac {q\lambda ^2-q^{-1}\lambda _r^2}{\lambda ^2-\lambda
  _r^2}q^{-\frac {1}{2}}(1-\Delta ^2\lambda ^{-2}q^{-1})^{N/4} \, .
\label{limRT}
\end{eqnarray}
The parameters $\lambda _r$ contained in this relation must satisfy a system of
Bethe equations which is obtained from (\ref {Bethe2}) by rescaling $\lambda
_r \rightarrow \lambda _r {\mu}^{1/2}$ and by taking the limit $\mu
\rightarrow 0$. The equations obtained in such a way are the Bethe equations
(\ref{rbethel2}) for the right conformal theory, in which $N$ is replaced by
$N/2$. Therefore, the r.h.s. of (\ref{limRT}) as function of $\la$ is
proportional by the factor $q^{-N/8}$ to the right conformal eigenvalue
(\ref{rT}), in which $N$ is replaced by $N/2$:
\ba
\lim _{\mu \rightarrow 0}
{\bf \Lambda}(\lambda \mu ^{1/2}, \{ \lambda_r \mu^{1/2} \}) &=& q^{l}\prod
_{r=1}^l \frac{q^{-1}\lambda^2-q\lambda _r^2}{\lambda ^2-\lambda_r^2}
q^{-\frac{1}{2}} (1-\Delta^2 \lambda^{-2}q)^{N/4}+ \nonumber \\
&+& q^{-l}\prod_{r=1}^l \frac{q\lambda^2-q^{-1}\lambda_r^2}{\lambda^2-\lambda_r^2}
q^{-\frac{1}{2}}(1-\Delta^2\lambda^{-2}q^{-1})^{N/4} = \nn \\
&=& q^{-N/8}  \{ q^{l}\prod _{r=1}^l  \frac {q^{-1}\lambda^2-q\lambda
_r^2}{\lambda^2-\lambda_r^2} q^{\frac{1}{2} \left(\frac{N}{4}-1 \right)}
(1-\Delta^2 \lambda^{-2} q )^{N/4}+ \nn \\
&+& q^{-l}\prod _{r=1}^l
\frac{q\lambda^2-q^{-1}\lambda_r^2}{\lambda^2-
\lambda_r^2}q^{\frac{1}{2}\left( \frac{N}{4} - 1 \right) }
(1-\Delta^2\lambda^{-2} q^{-1} )^{N/4}  \}  \, .
\ea

\medskip

Let us now consider the eigenvalue (\ref{LT}) and let us perform the following
limit:
\ba
\lim _{\mu \rightarrow 0}
{\bf \Lambda}^\prime (\lambda \mu ^{-1/2}, \{ \lambda _r \mu ^{-1/2} \} )&=&
q^{-l}\prod _{r=1}^l
\frac {q^{-1}\lambda
  ^2-q\lambda _r^2}{\lambda ^2-\lambda
  _r^2}q^{\frac {1}{2}}(1-\Delta ^2 \lambda ^2q^{-1})^{N/4}+\nonumber \\
 &+&q^{l}\prod _{r=1}^l
\frac {q\lambda ^2-q^{-1}\lambda _r^2}{\lambda ^2-\lambda
  _r^2}q^{\frac {1}{2}}(1-\Delta ^2\lambda ^2q)^{N/4} \, .
\label{Tright}
\ea

The parameters $\lambda _r$ contained in this relation must satisfy a system of Bethe
equations which is obtained from (\ref{Bethe4}) by rescaling $\lambda_r$ into
$\lambda _r {\mu}^{-1/2}$ and by taking the limit $\mu \rightarrow
0$. These equations are the Bethe equations for left conformal
theories (\ref{bethel2}), in which $N$ is replaced by $N/2$. Hence, the
the r.h.s. in (\ref{Tright}) is proportional, by a factor $q^{N/8}$, to
the left conformal eigenvalue (\ref{T}) with $N$ replaced by $N/2$:
\ba
\lim _{\mu \rightarrow 0}
{\bf \Lambda}^\prime (\lambda\mu ^{-1/2}, \{\lambda_r
\mu^{-1/2}\})&=&q^{-l}\prod _{r=1}^l \frac {q^{-1}\lambda^2-q\lambda
_r^2}{\lambda ^2-\lambda   _r^2}q^{\frac {1}{2}}(1-\Delta ^2 \lambda
^2q^{-1})^{N/4}+\nonumber \\  &+&q^{l}\prod _{r=1}^l \frac {q\lambda
^2-q^{-1}\lambda _r^2}{\lambda ^2-\lambda   _r^2}q^{\frac {1}{2}}(1-\Delta
^2\lambda ^2q)^{N/4} = \nonumber \\ &=&q^{N/8}  \{ q^{-l}\prod _{r=1}^l
\frac {q^{-1}\lambda ^2-q\lambda _r^2}{\lambda ^2-\lambda _r^2}q^{-\frac {1}{2}
\left (\frac {N}{4}-1\right)}(1-\Delta ^2\lambda ^2q^{-1})^{N/4}+ \nn \\
&+&q^{l}\prod _{r=1}^l
\frac {q\lambda ^2-q^{-1}\lambda _r^2}{\lambda ^2-\lambda
  _r^2}q^{-\frac {1}{2}\left (\frac {N}{4}-1\right)}(1-\Delta ^2\lambda
^2q)^{N/4}  \}  \, .
\ea
\medskip

\section{Cylinder scaling limits.}
\setcounter{equation}{0}
In this Section we want to derive the scaling expressions for the
crical and off-critical monodromy matrices (\ref{leftmon}, \ref{rightmon},
\ref{Rightmon}, \ref{Leftmon}) in the cylinder limit defined by
\be
N\rightarrow \infty \spz {\mbox {and fixed}} \spz R\equiv N\Delta .
\label{csl}
\ee
The previous limit (\ref{csl}) will be taken in a rigorous way in a
forthcoming paper \cite{35} defining in this way the continuum cylinder limit,
whereas now we illustrate here a heuristic operatorial limit to gain further
clues about the physical meaning of the monodromy matrices  previously
analysed. However, we believe that the results we will show are substantially
correct \cite{35}.
\medskip

From the definitions of $V_m^\pm$ (\ref{Ccorr1},\ref{Ccorr2}) one obtains
immediately that their behavior in the cylinder scaling limit is
\begin{eqnarray}
&V^-_m= -\Delta \phi ^\prime (y_{2m})+O(\Delta ^2)&\, ,\label {lim-}\\
&V^+_m= -\Delta \bar \phi ^\prime
(\bar y_{2m})+O(\Delta ^2) & \, ,  \label {lim+}
\end{eqnarray}
where $y_{2m}=\bar y_{2m}=m\frac {R}{N}$.
Hence, in this limit the Lax
operators (\ref{lax}) behave as follows
\begin{equation}
L_{m}(\lambda)=1+\Delta {\cal
  L}\left (m\frac {R}{N},\lambda\right) +O(\Delta ^2) \quad, \quad
\bar L_{m}(\lambda ^{-1})=
1+\Delta \bar {\cal L}\left (m\frac {R}{N},\lambda ^{-1}\right)+O(\Delta ^2)
 \, ,
\label{scallax}
\end{equation}
where we have defined
\begin{equation}
{\cal L}(y,\lambda)\equiv\left ( \begin{array}{cc} i\phi^\prime (y)& \lambda
\\ \lambda & -i\phi ^\prime (y)\\ \end{array} \right )
\, , \quad \bar {\cal L}(\bar y,\lambda ^{-1})\equiv\left ( \begin{array}{cc}
i\bar \phi ^\prime (\bar y)& \lambda ^{-1} \\ \lambda ^{-1} & -i\bar \phi
^\prime (\bar y) \\ \end{array} \right ) \, .
\label{limlax}
\end{equation}
Finally, by using (\ref{scallax}) we have that the left (\ref{leftmon}) and
right (\ref{rightmon}) monodromy matrices assume in the cylinder scaling limit
the form
\begin{eqnarray}
&&M(\lambda)= \prod _{k=1}^{\stackrel
{N}{\leftarrow}}\left [1+\Delta {\cal L}\left ( k\frac {R}{N},\lambda\right
)+O(\Delta ^2)\right ] \rightarrow {\cal P}{\mbox {exp}}\int _0^Rdy \, {\cal
L}\left (y, \lambda \right ) \, , \label{lecon} \\
&&\bar M (\lambda)= \prod
_{k=1}^{\stackrel {N}{\leftarrow}} \left [1+\Delta
\bar {\cal L}\left ( k\frac {R}{N},\lambda ^{-1}\right )+O(\Delta ^2)\right ] \rightarrow
{\cal P}{\mbox {exp}}\int _0^Rd\bar y \,
\bar {\cal L} \left (\bar y, \lambda ^{-1} \right )\, .
\label{ricon}
\end{eqnarray}
\medskip

At this point it is important to observe the slight difference between the
limit expressions (\ref{lecon}), (\ref{ricon}) and the chiral and anti-chiral
monodromy matrices proposed in \cite{BLZ}. Indeed, writing formul\ae \,
(\ref{limlax}) in the following way  \begin{eqnarray}
&{\cal L}(y,\lambda)=i\phi ^\prime (y)H+\lambda (E+F)& \, , \nonumber \\
&\bar {\cal L}(y,\lambda )=i\bar \phi ^\prime (y)H+\lambda (E+F)& \, ,
\end{eqnarray}
where
\begin{equation}
H=\left ( \begin{array}{cc} 1 & 0 \\ 0 & -1 \\ \end{array} \right )
\quad , \quad E=\left ( \begin{array}{cc} 0 & 1 \\ 0 & 0 \\ \end{array}\right )
\quad , \quad F=\left ( \begin{array}{cc} 0 & 0 \\ 1 & 0 \\
\end{array}\right )\, ,
\end{equation}
we finally obtain these expressions for (\ref{lecon}) and (\ref{ricon})
respectively:
\be
M(\lambda )={\cal P}{\mbox {exp}}\int _0^Rdy \left[
i\phi^\prime(y) H+ \lambda (E+F) \right]
\, ,
\ee
and
\be
\bar M (\lambda )={\cal P}{\mbox {exp}}\int _0^R d{\bar y} \left[
i\bar \phi^{\prime}(\bar y)H + \lambda ^{-1}(E+F) \right]
\, .
\ee
We will show in a forthcoming article \cite{35} how to
reproduce, starting from a regularized expression on a lattice, the
chiral and anti-chiral monodromy matrices of \cite{BLZ} and why these verify
the Yang-Baxter algebra instead of our braided version.
\medskip

Let us now derive the expressions for the monodromy matrices
(\ref {Rightmon}-\ref {Leftmon}) in the cylinder scaling limit. For what
concerns the monodromy matrix (\ref {Rightmon}) we have:
\begin{eqnarray}
{\bf M}(\lambda)&= &\prod _{i=1}^{\stackrel {N/4}{\leftarrow}}\left [1+\Delta \bar
{\cal L}\left (\frac {4i}{N}R, \frac {\mu ^{1/2}}{\lambda }\right )+O(\Delta ^2)\right ]
\left [1+\Delta \bar
{\cal L}\left (\frac {4i-1}{N}R, \frac {\mu ^{1/2}}{\lambda }\right )+O(\Delta ^2) \right ]
\cdot \nonumber \\
&\cdot & \left [1+\Delta
{\cal L}\left (\frac {4i-2}{N}R, {\mu ^{1/2}}{\lambda }\right )+O(\Delta ^2) \right ]
\left [1+\Delta
{\cal L}\left (\frac {4i-3}{N}R, {\mu ^{1/2}}{\lambda }\right )+O(\Delta ^2)\right ]
\nonumber \\
&=&\prod _{i=1}^{\stackrel {N/4}{\leftarrow}}\left [1+\Delta \bar
{\cal L}\left (\frac {4i}{N}R, \frac {\mu ^{1/2}}{\lambda }\right )
+\Delta \bar
{\cal L}\left (\frac {4i-1}{N}R, \frac {\mu ^{1/2}}{\lambda }\right
)+ \right. \nonumber \\
&+&\left. \Delta
{\cal L}\left (\frac {4i-2}{N}R, {\mu ^{1/2}}{\lambda }\right )
+\Delta
{\cal L}\left (\frac {4i-3}{N}R, {\mu ^{1/2}}{\lambda }\right )+O(\Delta ^2)\right]
\rightarrow
\nonumber \\
&\rightarrow&{\cal P}{\mbox {exp}}\frac {1}{2}\int _0^Rdy\left [ \bar
{\cal L}\left (y, \frac {\mu ^{1/2}}{\lambda }\right )
+{\cal L}\left (y, {\mu ^{1/2}}{\lambda }\right)\right ] \equiv {\cal M}(\lambda )  \, . \label{limmon}
\end{eqnarray}
In the last row we have defined the {\it scaling limit} monodromy matrix
${\cal M}(\lambda)$, because we find it again performing the limit (\ref{csl})
on (\ref {Leftmon}):
\begin{eqnarray} {\bf M}^\prime (\lambda)&=&\prod
_{i=1}^{\stackrel {N/4}{\leftarrow}}\left [1+\Delta {\cal L}\left (\frac
{4i}{N}R, \frac {\mu ^{1/2}}{\lambda }\right )+O(\Delta ^2)\right ] \left
[1+\Delta {\cal L}\left (\frac {4i-1}{N}R, \frac {\mu ^{1/2}}{\lambda }\right
)+O(\Delta ^2) \right ] \cdot \nonumber \\
&\cdot & \left [1+\Delta \bar
{\cal L}\left (\frac {4i-2}{N}R, {\mu ^{1/2}}{\lambda }\right )+O(\Delta ^2) \right ]
\left [1+\Delta \bar
{\cal L}\left (\frac {4i-3}{N}R, {\mu ^{1/2}}{\lambda }\right )+O(\Delta ^2)\right ]\nonumber \\
&=&\prod _{i=1}^{\stackrel {N/4}{\leftarrow}}\left [1+\Delta
{\cal L}\left (\frac {4i}{N}R, \frac {\mu ^{1/2}}{\lambda }\right )
+\Delta
{\cal L}\left (\frac {4i-1}{N}R, \frac {\mu ^{1/2}}{\lambda }\right
)+ \right. \nonumber \\
&+&\left. \Delta \bar
{\cal L}\left (\frac {4i-2}{N}R, {\mu ^{1/2}}{\lambda }\right )
+\Delta \bar
{\cal L}\left (\frac {4i-3}{N}R, {\mu ^{1/2}}{\lambda }\right )+O(\Delta ^2)
\right]\rightarrow \nonumber \\
&\rightarrow&{\cal P}{\mbox {exp}}\frac {1}{2}\int _0^Rdy\left [ \bar
{\cal L}\left (y, \frac {\mu ^{1/2}}{\lambda }\right )
+{\cal L}\left (y, {\mu ^{1/2}}{\lambda }\right)\right ] = {\cal M}(\la)  \, .
\label{limmon2}
\end{eqnarray}

On the basis of this coincidence we guess the {\it equivalence} of the
theories described by the two off-critical monodromy matrices in the
continuum  cylinder limit. Combining these heuristic results with the previous
ones, we can better support our conjecture according to which the monodromy
matrices (\ref{Rightmon}) and  (\ref {Leftmon}) are {\it equivalent}
descriptions of minimal conformal theories perturbed by the primary operator
$\Phi_{1,3}$.

\section{Similarity with Lattice Sine-Gordon Theory.}
\setcounter{equation}{0}
The interpretation of monodromy matrices ${\bf M}$ and ${\bf M}^\prime$ as
lattice regularized descriptions of $\Phi_{1,3}$ perturbation of CFT's will be
reinforced by the results of this Section. Indeed, we will show that BE's and
transfer matrices eigenvalues, derived for ${\bf M}$ and ${\bf M}^\prime$,
are strictly related to those of Lattice Sine-Gordon Theory (LSGT). In its
turn the continuum Sine-Gordon Theory (ST) contains the minimal
CFT's perturbed by $\Phi _{1,3}$ as a sub-theory derived through quantum
group reduction \cite{BLRS}.

The continuum ST on a cylinder is defined by the hamiltonian
\begin{equation}
H=\int _0^R dx \left [ \frac {1}{2} (\partial _t \Phi )^2 + \frac {1}{2}
(\partial _x \Phi )^2 + \frac {m^2}{8\gamma}(1-\cos {\sqrt {8\gamma}}
\Phi )\right ] \, ,
\label{Hamilt}
\end{equation}
where $m$ is the mass parameter and $\gamma$ the coupling constant. In the
paper \cite{IK} the authors found a lattice regularization of the ST
(\ref{Hamilt}) and hence they wrote the Bethe Equations and the eigenvalues of
the transfer matrix. With the definition
\be
S\equiv \left(\frac {1}{4}m\Delta\right)^2,
\ee
and for $N/4 \in {\Bbb N}$ these can be written as:

\begin{itemize}

\item  Bethe Equations
\begin{equation}
\left [ \frac {1+S({\lambda _s^\prime}^2e^{-i\gamma}+{\lambda _s^\prime}^{-2}e^{i\gamma})}
{1+S({\lambda _s^\prime}^{-2}e^{-i\gamma}+{\lambda _s^\prime}^{2}e^{i\gamma})} \right]^{N/4}=\prod_{\stackrel {r=1}{r\not= s}}^l \frac {{\lambda _r^\prime}^2e^{-i\gamma}-{\lambda _s^\prime}^2e^{i\gamma}}{{\lambda _r^\prime}^2e^{i\gamma}-{\lambda _s^\prime
}^2e^{-i\gamma}} \, ,
\end{equation}

or, after defining $\lambda_r^\prime=e^{\alpha _r^\prime }$,
\begin{equation}
\left [ \frac {1+2S\cosh (2\alpha^\prime _s-i\gamma)}
{1+2S\cosh (2\alpha _s^\prime +i\gamma)}
\right]^{N/4}=\prod_{\stackrel {r=1}{r\not= s}}^l \frac {\sinh (\alpha _s^\prime
-\alpha _r^\prime +i\gamma)} {\sinh (\alpha _s^\prime  -\alpha _r^\prime
-i\gamma)} \, ;
\label{SGBE}
\end{equation}

\item Eigenvalues of the transfer matrix
\begin{eqnarray}
\Lambda ^{IK}(\lambda ^\prime, \{\lambda _r^\prime\})&=&\prod _{r=1}^l
\frac {{\lambda _r^\prime}^2e^{i\gamma}-{\lambda
^\prime}^2e^{-i\gamma}}{{\lambda _r^\prime}^2-{\lambda ^\prime}^2} [
1+S({\lambda ^\prime}^2e^{-i\gamma}+{\lambda ^\prime}^{-2}e^{i\gamma}
)]^{N/4}+\nonumber \\ &+&\prod _{r=1}^l\frac {{\lambda
_r^\prime}^2e^{-i\gamma}-{\lambda ^\prime} ^2e^{i\gamma}}{{\lambda
_r^\prime}^2-{\lambda ^\prime}^2} [ 1+S({\lambda
^\prime}^2e^{i\gamma}+{\lambda ^\prime}^{-2}e^{-i\gamma})]^{N/4} \, ,
\end{eqnarray}

or, after defining $\lambda ^\prime =e^{\alpha ^\prime}, \lambda _r
^\prime =e^{\alpha _r^\prime }$,
\begin{eqnarray} \Lambda ^{IK}(\alpha^\prime
, \{\alpha _r^\prime \})&=&\prod _{r=1}^l\frac {\sinh (\alpha ^\prime -\alpha
_r^\prime -i\gamma)}{\sinh (\alpha^\prime -\alpha _r^\prime )} [ 1+2S\cosh
(2\alpha^\prime  -i\gamma)]^{N/4}+\nonumber \\ &+&\prod _{r=1}^l\frac  {\sinh
(\alpha^\prime  -\alpha  _r^\prime +i\gamma)}{\sinh (\alpha^\prime -\alpha
_r^\prime )} [ 1+2S\cosh (2\alpha ^\prime +i\gamma)]^{N/4} \, .
\label{SGE}
\end{eqnarray}

\end{itemize}

If we start from our trigonometric Bethe Equations  (\ref {Bethe3}, \ref
{Bethe5}) and eigenvalues (\ref{RTtrig}, \ref{LTtrig}) of the two transfer
matrices in the off-critical case and make the identifications:
\begin{equation}
\beta ^2 =\frac {\gamma}{\pi} \quad , \quad \frac
{e^{-2\Theta}}{1+e^{-4\Theta}}=S \quad , \quad \alpha = \alpha ^\prime +\frac
{i\pi}{2} \quad , \quad \alpha _r = \alpha ^\prime _r +\frac {i\pi}{2} \, ,
\label{identi}
\end{equation}
we then see that our Bethe Equations are equal to Sine-Gordon ones up to the
factors $e^{\mp 2i\pi\beta^2l}$. And, in addition, our eigenvalues
of ${\bf T}$ and ${\bf T}^\prime$ are proportional by the factor $e^{\pm \frac
{i\pi\beta^2}{2}}\left (\frac {1+e^{-4\Theta}}{4}\right)^{N/4}$ to Sine-Gordon
eigenvalues (\ref{SGE}), but the first addend has been multiplied by the factor
$e^{\pm i\pi\beta^2l}$ and the second by the factor $e^{\mp i\pi\beta^2l}$.
The upper sign in the exponentials ({\it the twist factors}) is for Bethe
states diagonalizing ${\bf T}$, the lower sign for Bethe states diagonalizing
${\bf T}^\prime$: the states which diagonalize ${\bf T}$ give rise to Bethe
equations with twist $e^{- 2i\pi\beta^2l}$, the states which diagonalize ${\bf
T}^\prime$ give rise to Bethe equations with twist $e^{+2i\pi\beta^2l}$.

Twisted versions of Bethe equations and eigenvalues of the transfer matrix for the
Sine-Gordon model are already present in the literature. However, usually the
twist is introduced {\it ad hoc} \cite{FMQR}, in order to identify the
properties of the states under the symmetry of the theory (\ref {Hamilt})
$\Phi \rightarrow \Phi +  \frac{2\pi n}{\sqrt{8\gamma}}$.  On the contrary, in
our case the dynamical twist comes naturally into the theory and, differently
from usual approaches also to other theories, it depends on the number $l$ of
Bethe roots.

For instance we want to show how we recover the $l$- and $N$-independent twist
introduced in \cite{FMQR} in the particular case $\beta ^2 =\frac {1}{p+1}$,
with $p$ positive integer. We call {\it the vacuum sector} solutions those
sets of Bethe roots corresponding to $l=N/4$ in the limit
$N\rightarrow\infty$. In this limit, we are obliged to parameterize the chain
length as follows (this kind of parameterization has been also used in
\cite{FTi}  in the case of the Liouville model)
\begin{equation} \frac
{N}{4}=(p+1)n+{\kappa} \quad , \quad 0\leq {\kappa}\leq p \quad , \quad n \in
{{\Bbb N}} \, .
\label {par}
\end{equation}
Indeed, at fixed $\kappa$ the twist phase factors do not oscillate as
$N\rightarrow \infty$
\begin{equation} e^{\mp 2i\pi\beta ^2 l}=e^{\mp {2i\pi}\frac {1}{p+1}\frac
{N}{4}}\rightarrow e^{\mp 2i\pi\frac {1}{p+1}{\kappa}} \, ,
\end{equation}
but become $N$-independent. Hence, for any $\kappa$, the Bethe equations
(\ref {Bethe3}, \ref {Bethe5}) and the corresponding Bethe state become in a
natural way respectively the Bethe equations and the the $\kappa$-vacuum of the
twisted Sine-Gordon model presented in \cite{FMQR}. Besides, for $\kappa\neq 0$
this $\kappa$-vacuum is also a state of the $p$-th unitary minimal CFT. This
procedure can be repeated also for excited states, which are characterized as
well as the vacuum by their twisting properties, and for non-unitary models. We
will come back to this point in a forthcoming publication \cite {35}. Of
course, for the non-twisted state ($\kappa =0$), we obtain the LSGT Bethe
Equations (\ref{SGBE}) for the vacuum and the corresponding eigenvalue of the
transfer matrix proportional to (\ref{SGE}).

\section{Conclusions e Perspectives.}
We have found a generalization of the Yang-Baxter algebra, called braided Yang-Baxter
algebra, as a result of discretization and quantization of the monodromy
matrices of two coupled (m)KdV equations. A matrix $Z_{ab}(q)$,
independent of the spectral parameter and of the lattice variables, encodes
the braiding effect, which is a pure quantum feature and disappears in the
classical limit $q\rightarrow 1$, because $Z_{ab}(q)\rightarrow \buno$. By
virtue of the commutativity of the braiding matrix $Z_{ab}$ with the quantum
$R$-matrix we have proved that the braided Yang-Baxter algebra still ensures
the Liouville integrability, {\it i.e.} that the transfer matrix commutes
for different values of the spectral parameter and therefore generates
(an infinite number of) operators in involution. Regarding these operators as a
Cartan sub-algebra, a suitable generalization of Algebraic Bethe Ansatz
technique has been built to construct representations in which they are
diagonal. As an effect of the braiding an $l$-dependent {\it dynamical} twist
appears in the Bethe Equations.

We will prove in a forthcoming paper \cite{35} that these representations are
vacuum (highest weight) representations for the hamiltonian operator. In the
cylinder continuum limit, we will find non-linear integral equations
describing the energy spectrum. The conjecture, we have here proposed and
supported, that this spectrum is that of Perturbed Minimal
Conformal Field Theory, will be there proved.

Actually, our left and right (conformal) monodromy matrices (\ref{leftmon}) and
(\ref{rightmon}) are in the cylinder continuum limit slightly different from
those analyzed in \cite{BLZ}, and it is very peculiar that they form a braided
Yang-Baxter algebra, although those in \cite{BLZ} close an usual Yang-Baxter
algebra. Nevertheless, we will see in a forthcoming paper \cite{35} how to
build, from our monodromy matrices, others satisfying the UN braided
Yang-Baxter relation \cite{PhD}, realizing a deeper link to \cite{BLZ}.

In a sequel of paper \cite{FS1, FS2, FS3}, one of the author (DF) in
collaboration with M. Stanishkov has built a general method of finding hidden
symmetries in the classical KdV theory starting from the Lax operator
(\ref{lleft}) of Section 2. In particular, a very interesting quasi-local
Virasoro algebra has been discovered in \cite{FS4, F} and its action on
soliton solutions has been studied. Since only some hints have been given
about quantization of this intriguing symmetry algebra, it is very interesting
to understand the arising of this algebra in the quantum context of the
present paper.

Eventually, this way of quantizing the simplest KdV theory and of going out
off-criticality grounds only on algebraic properties of the involved
fields/variables and consequently leads very easily to applications to all the
generalized KdV theories \cite{PhD}. Among them the next interesting case
would be represented by the quantum $A_2^{(2)}$ KdV depicted in \cite{FRS},
which completes the scenario of integrable perturbations of minimal Conformal
Field Theories ({\it i.e.} theories without extended conformal symmetry
algebra).

\medskip

{\bf Acknowledgments} - D.F. is particularly grateful to M. Stanishkov for
endless discussions and suggestions. We are also indebted to F. Colomo, A.
Cappelli, E. Corrigan, G. Delius, P.E. Dorey, A. Kundu, N. MacKay, G. Mussardo,
R. Weston, R. Tateo for discussions and interest in this work. D.F. thanks
EPRSC for the grant GR/M66370. M.R. thanks CNR/NATO for the grant 215.31 and
EPRSC for the grant GR/M97497. The work has been partially supported by EC TMR
Contract ERBFMRXCT960012.

\end{document}